\documentclass[twocolumn,superscriptaddress,aps]{revtex4-2}
\usepackage{amsmath}
\usepackage{amssymb}
\usepackage{graphicx}
\graphicspath{{./Figures/}}
\usepackage{color}
\usepackage[colorlinks=true,linkcolor=blue,citecolor=blue]{hyperref}

\usepackage{siunitx}
\usepackage{amsfonts}
\usepackage{bm}
\usepackage{euscript}
\usepackage{textcomp}
\usepackage{gensymb}
\usepackage{float}
\usepackage{enumitem}
\usepackage{xcolor}
\usepackage{mhchem}

\usepackage{multibib}
\usepackage{bbm}

\newcommand{\blutx}[1]{\textcolor{blue}{}}

\newcommand{\xqwtb}{\chi^{+-}_{0}\left( {\bf q}, \omega \right)}

\newcommand{\ispone}{\downarrow}
\newcommand{\isptwo}{\uparrow}
\newcommand{\bfR}{{\bf R}}
\newcommand{\bfr}{{\bf r}}
\newcommand{\bfq}{{\bf q}}
\newcommand{\bfk}{{\bf k}}

\begin{document}

\title{Chiral and flat-band magnetic cluster excitations in a ferromagnetic kagome metal}

\author{S. X. M. Riberolles}
\affiliation{Ames Laboratory, Ames, IA, 50011, USA}

\author{Tyler J. Slade}
\affiliation{Ames Laboratory, Ames, IA, 50011, USA}

\author{Tianxiong Han}
\affiliation{Ames Laboratory, Ames, IA, 50011, USA}
\affiliation{Department of Physics and Astronomy, Iowa State University, Ames, IA, 50011, USA}

\author{Bing Li}
\affiliation{Ames Laboratory, Ames, IA, 50011, USA}
\affiliation{Department of Physics and Astronomy, Iowa State University, Ames, IA, 50011, USA}

\author {D.~L.~Abernathy}
\affiliation{Oak Ridge National Laboratory, Oak Ridge, TN 37831 USA}

\author{P.~C.~Canfield}
\affiliation{Ames Laboratory, Ames, IA, 50011, USA}
\affiliation{Department of Physics and Astronomy, Iowa State University, Ames, IA, 50011, USA}

\author{B. G. Ueland}
\affiliation{Ames Laboratory, Ames, IA, 50011, USA}

\author{P. P. Orth}
\affiliation{Ames Laboratory, Ames, IA, 50011, USA}
\affiliation{Department of Physics and Astronomy, Iowa State University, Ames, IA, 50011, USA}

\author{Liqin Ke}
\affiliation{Ames Laboratory, Ames, IA, 50011, USA}

\author{R.~J.~McQueeney}
\affiliation{Ames Laboratory, Ames, IA, 50011, USA}
\affiliation{Department of Physics and Astronomy, Iowa State University, Ames, IA, 50011, USA}

 \date{\today}

\begin{abstract}
TbMn$_{6}$Sn$_{6}$ is a metallic ferrimagnet that displays signatures of band topology arising from a combination of uniaxial ferromagnetism and spin-orbit coupling within its Mn kagome layers. Whereas the low energy magnetic excitations can be described as collective spin waves using a local moment Heisenberg model, sharply defined optical and flat-band collective magnon modes are not observed. In their place, we find overdamped chiral and flat-band spin correlations that are localized to hexagonal plaquettes within the kagome layer.   
\end{abstract}

\maketitle

\section{Introduction}
Flat electronic bands are susceptible to a variety of instabilities driven by electron-electron correlations and band filling.  In particular, nearest-neighbor hopping on a two-dimensional (2D) kagome lattice guarantees flat bands while also hosting Dirac band crossings and band touchings that impart a non-trivial topology.  Recent discoveries of superconductivity \cite{Ortiz20} and charge ordering \cite{Jiang21} in $A$V$_3$Sb$_5$, and itinerant ferromagnetism (FM) with large (topological) anomalous Hall response in Co$_3$Sn$_2$S$_2$ \cite{Liu18} have elevated interest in kagome metals as an adaptable system to study the interplay of topology, superconductivity, magnetism, and other charge instabilities \cite{Yin22}.  

$R$Mn$_6$Sn$_6$ (where $R$ is a rare-earth) materials comprise an interesting class of magnetic kagome metals \cite{Ghimire20,Yin20,Ma21,Li21,Dhakal21,Kabir22}.  The metallic Mn kagome layers have robust itinerant FM order that can be manipulated by interleaved FM $R$ triangular layers through $R$--Mn antiferromagnetic coupling.  In TbMn$_6$Sn$_6$, $R$--Mn coupling and the uniaxial anisotropy of the Tb ions forces Mn moments to orient perpendicular to the kagome layer \cite{Idrissi91}, creating an ideal scenario for a Chern insulator where spin-orbit coupling (SOC) gaps out spin-polarized Dirac band crossings \cite{Yin20}.  In $R$Mn$_6$Sn$_6$ and other kagome metals with FM layers, such as FeSn \cite{Kakihana18, Kang20, Ye20, Han21} and Fe$_3$Sn$_2$ \cite{Kida11,Lin18,Ye19}, the connection between itinerant FM order and flat electronic bands, evidence of the magnetic coupling to Dirac fermions, and the role of magnetic fluctuations in topological phenomena are all open questions.

These key questions can be addressed using inelastic neutron scattering (INS) measurements to probe magnetic excitations and determine their coupling to kagome electronic bands. The high energy of flat and optical magnon bands in FM kagome lattices, which also host topological magnonic features \cite{Onose10,Mook14,Mook14_2}, make them susceptible to Landau damping by particle-hole excitations (Stoner continuum).  The heavy damping that obscures the observation of these modes in a variety of FM kagome metals supports this hypothesis \cite{Zhang20,Xie21,Do21,Riberolles22}. 

Here, we use INS measurements on TbMn$_6$Sn$_6$ to provide key insights into the high energy magnetic excitations of a FM kagome metal.  Whereas the low-energy acoustic magnon modes are well-defined collective excitations, we observe unusual excitations at high energies that are described by short-ranged spin correlations on a hexagonal plaquette in the kagome layer.  Excitations at the K-point of the Brillouin zone consist of spin correlations on a hexagonal plaquette that are derived from the expected Wannier states associated with a flat magnon mode \cite{Bergman08}.  Excitations at the zone center ($\Gamma'$) exhibit chiral antiferromagnetic correlations on the hexagonal plaquette that are commonly associated with magnetic frustration \cite{Inami00, Grohol05}.  Density-functional theory (DFT) calculations assess that the overdamped character of these modes can originate from Landau damping \cite{Fawcett88, Diallo09, Chen20} which is not likely to arise from massive Dirac fermions lying close to the Fermi energy \cite{Yin20,Ma21,Li21,Liu21,Gu22}.  This discovery of chiral magnetic cluster excitations raises new questions about their itinerant character and their coupling to other novel electronic features of the kagome lattice, such as orbital plaquette currents \cite{Mielke22}.




 \section{Experimental Details}
Single crystals of Tb166 were grown from excess Sn using the flux method as previously described \cite{Riberolles22}. INS measurements were performed on the Wide Angular-Range Chopper Spectrometer (ARCS) at the Spallation Neutron Source at Oak Ridge National Laboratory. An array of nine crystals with a total mass of 2.56 grams was co-aligned with the ($H$,0,$L$) scattering plane set horizontally, and attached to the cold head of a closed-cycle-refrigerator. The data were collected at the base temperature of 5 K using incident energies of $E_i =$ 250 and 500 meV.  For each $E_i$ measurement, the sample was rotated around the vertical axis to increase the \textbf{q} coverage. The neutron scattering data are described using the momentum transfer in hexagonal reciprocal lattice units, ${\bf q}(H,K,L) = \frac{2\pi}{a}\frac{2}{\sqrt{3}}(H\hat{a}^*+K\hat{b}^*)+\frac{2\pi}{c}L\hat{z}$. The INS data are presented in terms of the orthogonal vectors $(1,0,0)$ and $(-1,2,0)$, as shown in Fig.~\ref{dispersion}(b). We describe the data with reference to special points in the 2D Brillouin zone; $\Gamma$--(0,0), M--($\frac{1}{2}$,0), and K--($\frac{1}{3}$,$\frac{1}{3}$). The INS data are displayed as intensities that are proportional to the spin-spin correlation function $S({\bf q},E)$, where $E$ is the energy. To improve statistics, the data have been symmetrized with respect to the crystallographic space group P6/$mmm$.

\begin{figure}
\includegraphics[width=1.0\linewidth]{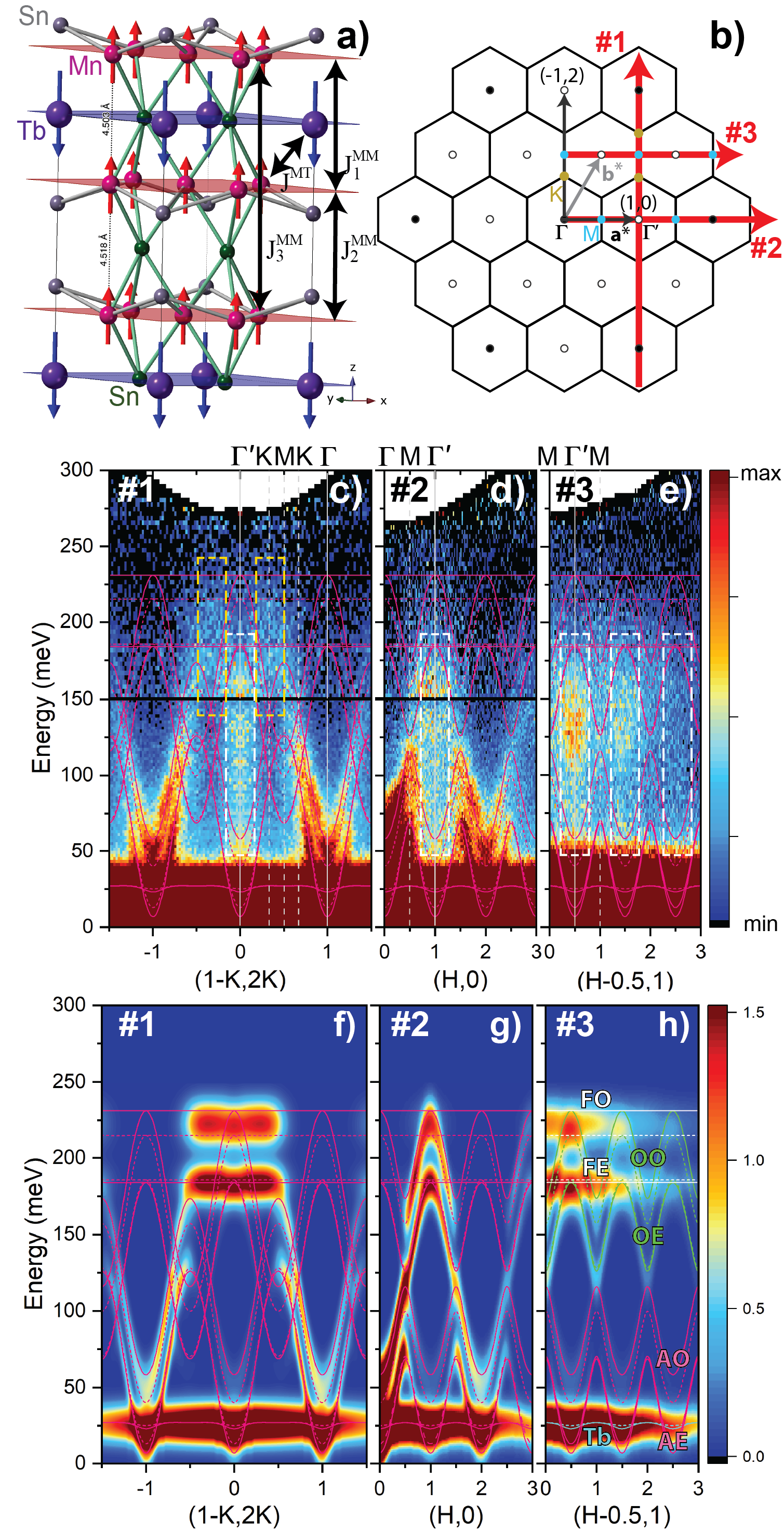}
\caption{\footnotesize (a) Ferrimagnetic structure of Tb166 with key interlayer magnetic interactions indicated by black arrows.  (b) 2D hexagonal Brillouin zone showing reciprocal lattice vectors $\textbf{a}^{*}=(1,0)$ and $\textbf{b}^{*}=(0,1)$ and representative high symmetry $\Gamma$ (black circles), $\Gamma'$ (empty circles), M (blue circles), and K (gold circles) points. INS data are described with the orthogonal vectors (1,0) and (-1,2). Red arrows correspond to the three reciprocal space slices shown in panels (c)--(h). Panels (c)--(e) show the intensity of slices through the INS data along the $(1-K,2K)$ (slice \#1), $(H,0)$ (slice \#2), and $(H-0.5,1)$ (slice \#3) directions, respectively, after averaging over $L=0-7$. In panels (c) and (d), data with $E<150$ meV ($> 150$ meV) were collected with $E_i=250$ meV (500 meV), respectively. Data in (e) were collected with $E_i=500$ meV. Panels (f)-(h) show identical slices as (c)--(e) calculated from linear spin wave theory using the parameters described in the main text.  In panels (c)--(g), the solid and dashed pink lines correspond to model dispersions with $L=0$ and $L=1/2$, respectively.  In panel (h), blue, pink, green, and white lines label the model dispersions of the Tb, AE/AO, OE/OO, and FE/FO modes, respectively, as described in the text.  White (yellow) dashed rectangles outline overdamped cluster excitations at the $\Gamma'$~(K) point, respectively.}
\label{dispersion}
\end{figure} 

\section{INS data}
The magnetic sublattice of TbMn$_6$Sn$_6$ consists of stacked Mn kagome and Tb triangular layers, as shown in Fig.~\ref{dispersion}(a).  The FM Mn and Tb layers couple antiferromagnetically, resulting in a ferrimagnetic ground state with moments pointing along the $c$-axis.  Within linear spin wave theory, TbMn$_6$Sn$_6$ has seven spin wave branches. We label these branches as acoustic--even (AE), acoustic--odd (AO), optical--even (OE), optical--odd (OO), flat--even (FE), flat--odd (FO), and Tb.  As there are two kagome layers in the unit cell, the even (odd) branches correspond to in-phase (out-of-phase) precession of Mn moments in adjacent kagome layers and have strong neutron intensity in Brillouin zones with $L=even$ ($L=odd$), respectively. The acoustic and optical branches are found below and above the K-point Dirac magnon crossing, respectively, and represent in-phase and out-of-phase precession of the three Mn moments within the unit cell of a single kagome layer. AE and AO branches have strong neutron intensity in $\Gamma$ zones with $H=even$ and $K=even$, whereas OE, OO, FE, and FO branches are strongest in $\Gamma'$ zones with $H=odd$ and $K=odd$ or $H+K=odd$, as shown in Fig.~\ref{dispersion}(b).

\begin{figure*}
\includegraphics[width=1.\linewidth]{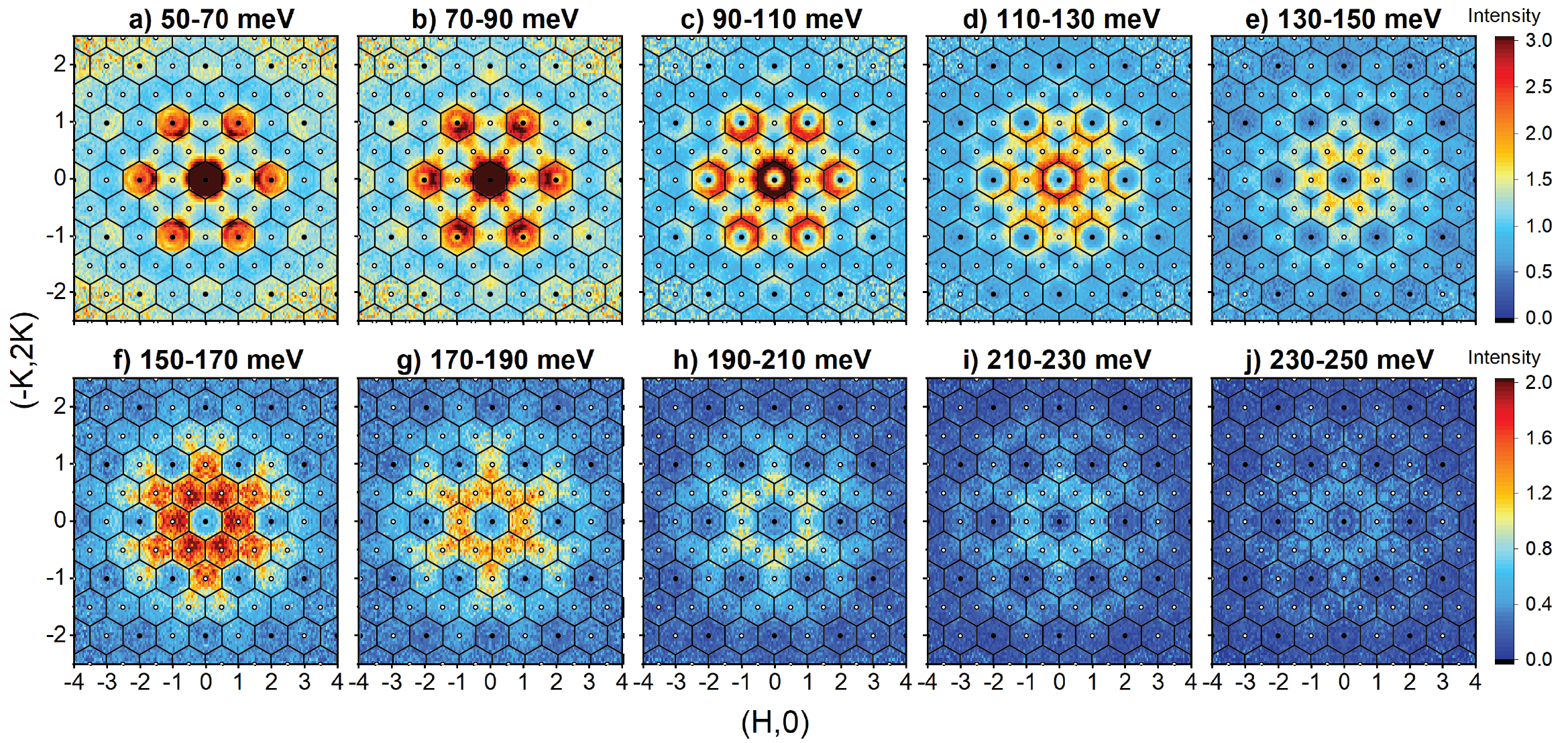}
\caption{\footnotesize Constant energy slices of the TbMn$_6$Sn$_6$ spin excitations over incremental energy ranges from (a) 50--70 meV, (b) 70--90 meV, (c) 90--110 meV, (d) 110--130 meV, (e) 130--150 meV, (f) 150--170 meV, (g) 170--190 meV, (h) 190--210 meV, (i) 210--230 meV, (j) 230--250 meV.  Data were collected at $E_i=250$ meV for panels (a)--(e) and $E_i=500$ meV for panels (f)--(j). All cuts are averaged over an $L$-range from -7 to 7 rlu. Hexagonal Brillouin zone boundaries are shown and $\Gamma$ and $\Gamma'$ zone centers are indicated by filled and empty circles, respectively.}
\label{constE}
\end{figure*} 

Previous INS experiments were conducted on smaller samples and mapped out the lowest-lying AE, AO and Tb branches below 125 meV \cite{Riberolles22}.  These spin waves possess sharp, dispersive excitations throughout the Brillouin zone, indicative of their collective nature, and are well represented by a Heisenberg model consisting of intralayer and interlayer pairwise exchange interactions and single-ion anisotropy terms, as described in the Supplementary Material (SM) \cite{SM}.  

In Ref.~\cite{Riberolles22}, we were unable to clearly observe the OE, OO, FE, and FO branches due to the small sample volume and increasingly broad line shapes encountered at higher energies. The measurements reported here were performed with a larger sample volume and higher incident energies, revealing significant magnetic spectral weight up to 250 meV that accounts for the missing branches in the previous data. However, as previous reports hinted, these higher energy features are {\it incoherent} (broad in both momentum and energy), unlike the collective nature of the AO, AE, and Tb branches.

Figures \ref{dispersion}(c)--(e) show slices of the data along different reciprocal space directions within the kagome layers, as indicated in Fig.~\ref{dispersion}(b). The data are averaged over $L$ to improve statistics, resulting in the simultaneous observation of even and odd branches. The data are compared to the Heisenberg model dispersions shown as pink lines. We also compare the data to model calculations of the INS intensities under the same reciprocal space averaging conditions, as shown in Fig.~\ref{dispersion}(f)--(h).  

In Figs.~\ref{dispersion}(c) and (d), slices \#1 and \#2 along the $(1-K,2K)$ and $(H,0)$ directions reveal dispersing AE and AO branches emanating from the $\Gamma$ points that highlight their collective character. In slice \#1, the AE and AO dispersions are well-defined up to their respective K-point Dirac crossings at 90 meV and 140 meV. In slice \#2, the AE and AO branches are well-defined up to the M-point with energies of 70 meV and 115 meV. The AE, AO, and Tb dispersions and intensities are consistent with model calculations shown in Fig. \ref{dispersion} (f)--(g).

Traces of broad OE, OO, FE and FO excitations can be seen as high as $\sim$ 250 meV in slices \#1--3, consistent with model predictions of a 230 meV energy cutoff.  However, these modes are  incoherent and overdamped. Incoherent excitations observed in slices \#1-3 form a steep feature centered in the $\Gamma'$ zones which extends from approximately 50 to 180 meV (white rectangle).  These modes are clearly observed in slice \#3, where the $\Gamma$-zone AO and AE modes are suppressed. Slice \#1 also indicates higher energy incoherent modes centered at the K-point and extending from 140--230 meV (yellow rectangle).

In an attempt to capture these features, the current Heisenberg model can be extended to include longer-range interactions within the kagome layer. These interactions can be added with constraints that fix the M-point AO and AE energies while attempting to lower the energy of the OO and OE branches at the $\Gamma'$-point. However, these interactions also introduce unsatisfactory distortions of the magnetic spectra (such as an overall lowering of the magnetic bandwidth) and the broad nature of the high-energy excitations present difficulties in numerical fitting of the extended models. More details of these extended models are provided in Fig.~S1 \cite{SM}.

Rather than pursuing an extended Heisenberg model, a qualitative understanding of the incoherent high-energy excitations can be obtained from constant energy cuts through the excitation spectra, as shown in Fig.~\ref{constE}.  Starting at 60 meV in Fig.~\ref{constE}(a), we see the intense AE and AO conical dispersions that form concentric rings of intensity in the $\Gamma$ zones.  We also see streaks of intensity that extend longitudinally across the $\Gamma'$ zones.  As we move up in energy, the AO branch eventually reaches the zone boundary around 115 meV [Fig.~\ref{constE}(d)] while the longitudinal streak in the $\Gamma'$ zone persists.  At energies above the AO branch cutoff, only the streak through the $\Gamma'$ zones remain [Fig.~\ref{constE}(e)].  As we continue to increase energy, the $\Gamma'$ streaks overlap with scattering intensity that forms at K-points found at the intersection between $\Gamma'$ zones [eg., at ${\bf q}=(\frac{2}{3},\frac{2}{3})$]. Figure~\ref{constE}(h) shows that only these K-point excitations remain at 200 meV.  All magnetic excitations disappear at about 250 meV. For comparison, Heisenberg model calculations of these constant-$E$ cuts can be found in Fig.~S2 \cite{SM}.  We can make several conclusions about these results; (1) the expected spectral weight of the OE/OO/FE/FO branches is associated with incoherent excitations in the $\Gamma'$ zones, (2) the steep incoherent excitation centered at $\Gamma'$ in Fig.~\ref{dispersion} forms the longitudinal streaks in the $\Gamma'$ zones in Fig.~\ref{constE}(c)--(e), and (3) the K-point excitations extend to the highest energies as observed in Fig.~\ref{dispersion}(c).

\begin{figure}
\includegraphics[width=1.\linewidth]{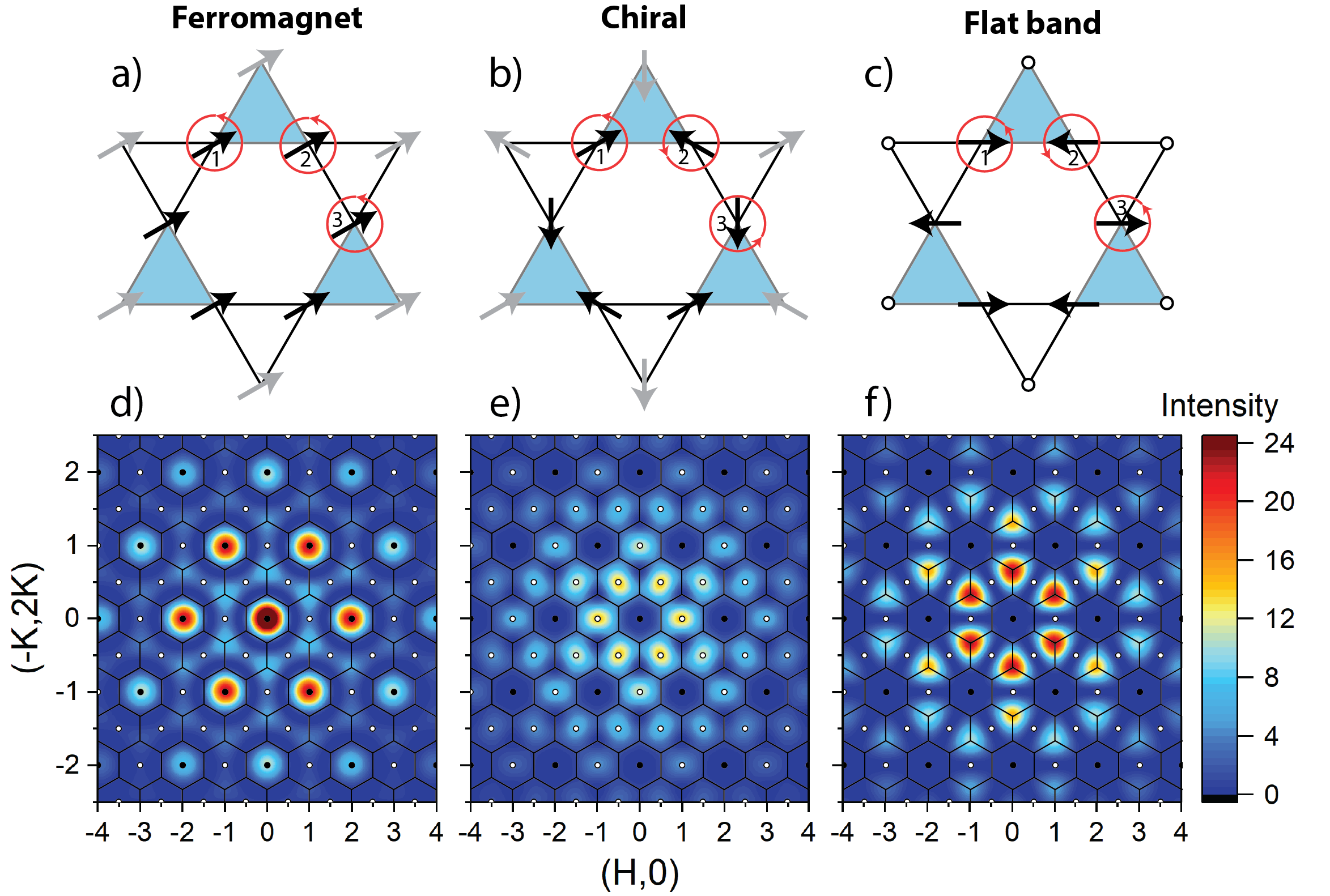}
\caption{\footnotesize Spin patterns for the in-plane component of the Mn moment in (a) ferromagnetic, (b) chiral, and (c) flat-band hexagonal clusters. Numbers label the unique spins in the kagome layer and red circles indicate the direction of spin precession. Panels (d)--(f) show the corresponding static structure factors for the black spins around the hexagonal plaquettes in (a)--(c), respectively.}
\label{cluster}
\end{figure}

The broad nature of the high energy modes in reciprocal space suggests that they are localized excitations formed from magnetic clusters within the kagome layer.  We studied both triangular and hexagonal plaquettes (see Fig.~S5 \cite{SM}) and find that the data is accurately described by spin clusters on a single hexagon, as shown in Figs.~\ref{cluster}(a)--(c).  Considering that the time-averaged Mn moments point perpendicular to the kagome layer, the spin patterns shown correspond to the instantaneous transverse components which precess around the $c$-axis. The neutron intensity is estimated by calculating the corresponding static structure factors
\begin{equation}
S({\bf q}) = f^2(q)\Big|\sum_{j=1,6} {\rm e}^{i\phi_j} {\rm e}^{i{\bf q}\cdot {\bf r}_j} \Big|^2
\end{equation}
 as shown in Figs.~\ref{cluster}(d)--(f).  Here, $f(q)$ is the magnetic form factor and $\phi_j$ describes the relative angle of the instantaneous spin direction of the transverse component of spin $j$. Dynamical spin precession averages over the spin direction at each site and only the relative angle between spins around the hexagon is relevant.

The FM cluster in Fig.~\ref{cluster}(a) consists of in-phase precession of the spins ($\phi_{j+1}=\phi_j$), resulting in strong intensity in the $\Gamma$ zones [Fig.~\ref{cluster}(d)]. This corresponds to the AO and AE modes, although these modes are more appropriately described as collective excitations using spin wave theory \cite{Riberolles22}.  The chiral cluster in Fig.~\ref{cluster}(b) is based on the $q=0$ antiferromagnet (AF), which is a chiral magnetic ground state found in kagome systems with nearest-neighbor AF interactions, such as iron jarosite \cite{Inami00, Grohol05}.  In the long-range-ordered $q=0$ AF state, magnetic Bragg peaks appear in $\Gamma'$ zones and are systematically absent in the $\Gamma$ zones.  In the chiral cluster, Mn spins successively rotate by 120$^\circ$ around the hexagon ($\phi_{j+1}=\phi_j + 2\pi/3$) (positive chirality) and result in an intensity pattern consistent with longitudinal streaks observed in the $\Gamma'$ zones [Fig.~\ref{cluster}(e)].  Positive and negative chiralities have the same intensity pattern and are degenerate in the absence of Dzyaloshinskii-Moriya interactions.  Finally, we consider a flat-band cluster, consisting of successive 180$^\circ$ rotations of spins around the hexagon ($\phi_{j+1}=\phi_j+\pi$) shown in Fig.~\ref{cluster}(c). This is the correct Wannier representation of the localized flat band excitations \cite{Bergman08} and generates scattering intensity which is peaked at the K-points surrounded by $\Gamma'$ zones [Fig.~\ref{cluster}(f)].  Thus, the two incoherent modes are associated with chiral and flat band spin correlations on isolated hexagonal plaquettes.  Damped harmonic oscillator analysis of the energy spectra (see Fig.~S3 \cite{SM}) indicate overdamped character with a quality-factor of 1.5--2.

\begin{figure}
\includegraphics[width=1.0\linewidth]{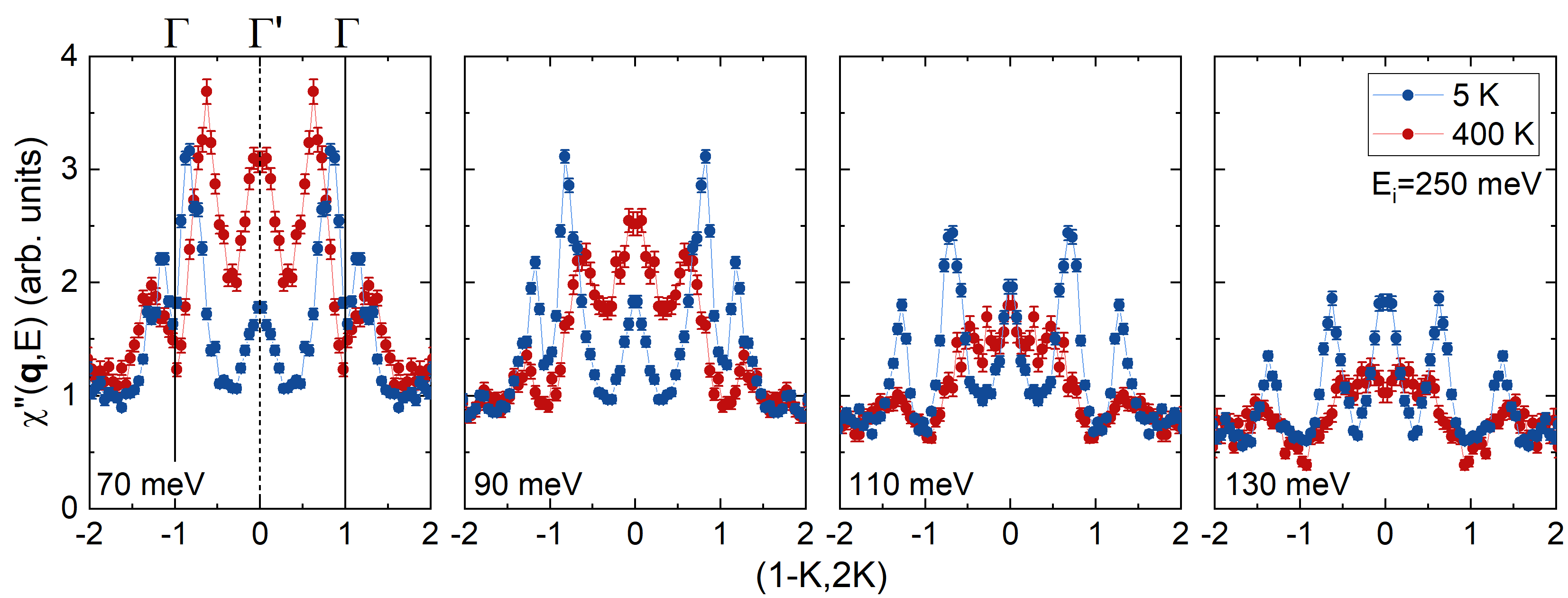}
\caption{\footnotesize Comparison of constant-energy cuts along the $(-K,2K)$ direction for $T=5$ K (blue) and 400 K (red).  Data are scaled to be proportional to the dynamical susceptibility, $\chi''({\bf q},E)=I({\bf q},E)(1-{\rm exp}(-E/k_{\rm B}T))$.  Cuts are performed with $E_i=250$ meV at energy transfers of 70, 90, 110, and 130 meV (averaged over an energy range of $\Delta E= \pm 5$ meV).  All plots are averaged over reciprocal space ranges of $H=$ [0.9:1.1] and $L=$ [-7:7]. }
\label{fig:temperature}
\end{figure}

Figure \ref{fig:temperature} and Fig.~S4 \cite{SM} compares cuts through excitations in the $\Gamma$ and $\Gamma'$ zones at $T=5$ K and 400 K, just below the Curie temperature of $T_c=420$ K.  Dispersive AO/AE modes and incoherent $\Gamma'$ chiral modes are both observed at 400 K. The intensity of both modes rapidly diminishes above 100 meV as a consequence of softening and mode damping near $T_c$.  The data suggest that the both the $\Gamma$ and $\Gamma'$ responses originate from the Mn sublattice magnetization.
  
\section{Discussion}
The uniaxial FM kagome layers in TbMn$_6$Sn$_6$ display conventional, collective acoustic spin waves when Mn spins precess in-phase within a kagome layer.  However, optical and flat band modes whose precessions are out-of-phase have an incoherent character, remaining localized to a single hexagonal plaquette with heavy damping.  Confinement to a hexagonal plaquette is expected for the flat band excitations of the kagome lattice due to phase cancellation of coherent spin precession on the triangular vertices surrounding a hexagon (see Fig.~S6 \cite{SM}).  Thus, the observed flat bands would, in principle, be consistent with linear spin wave theory were it not for the heavy damping.  Similar to their electronic counterparts, the observation of the flat magnon band in a FM metal is rare: Flat band magnon modes are only reported in the insulating kagome FM Cu[1,3-benzenedicarboxylate] where damping is small \cite{Chisnell15}, but have not previously been reported in FM kagome metals.

The optical modes are not expected to have a localized, chiral character based on the kagome geometry alone and are truly anomalous. Such localization could arise from magnetic frustration. For example, a similar confinement of excitations to hexagonal plaquettes is observed in frustrated pyrochlore antiferromagnets \cite{Lee02}.  Frustration from competing magnetic interactions are possible in FM kagome metals.  For example, itinerant FM order is proposed to coexist with $q=0$ chiral AF order in Co$_3$Sn$_2$S$_2$ \cite{Guguchia20, Lachman20, Zhang21}.  However, FM order is much more robust in $R$Mn$_6$Mn$_6$ and competing intralayer magnetic interactions are not expected to play a significant role.

Clues to this anomalous behavior could come from the itinerant-like character of the optical and flat magnon modes.  For example, the damping of both flat and optical/chiral modes observed here in TbMn$_6$Sn$_6$ and reported in other FM kagome metals, such as YMn$_6$Sn$_6$ \cite{Zhang20} and FeSn \cite{Do21, Xie21}.  Heavy damping is consistent with the coupling between magnons and electrons, providing a window into the correlated electron physics of kagome metals. In TbMn$_6$Sn$_6$, quantum oscillations, photoemission, and DFT calculations indicate massive Dirac fermions approximately 100 meV above $E_F$ \cite{Yin20,Ma21,Li21,Liu21,Gu22} which could provide decay channels for particle-hole excitations at the $\Gamma$-point (intravalley) and K-point (intervalley) (see in Fig.~S8 in the SM).  However, the massive Dirac fermions in TbMn$_6$Sn$_6$ are spin-polarized minority bands with an exchange splitting of $\sim1$ eV \cite{Lee22}, which should strongly suppress any spin-flip scattering channels. Our DFT calculations of the bare electronic spin-flip susceptibility agree with this assessment (Fig~S7 \cite{SM}).  However, the DFT calculations do point to potential sources of Landau damping from trivial electronic bands (see SM \cite{SM}).

In summary, we have observed the magnetic spectral weight associated with optical and flat magnon bands in a FM kagome metal. The excitations are comprised of chiral and flat band spin correlations within a hexagonal plaquette, but are incoherent between plaquettes. Extreme damping of these cluster excitations could be caused by Landau damping from trivial electronic bands, although the localized character of chiral modes may point to more exotic couplings. For example, future work should consider the possibility that staggered flux or orbital currents predicted to circulate on the same plaquettes \cite{Yin22,Mielke22} could impart chirality, localization, and damping to the magnetic excitations.

\section{Acknowledgments} RJM would like to thank Joe Checkelsky for comments. RJM, LK, PPO, BGU, BL, and SXMR's work at the Ames Laboratory is supported by the U.S. Department of Energy (USDOE), Office of Basic Energy Sciences, Division of Materials Sciences and Engineering. TJS, TH, and PC are supported by the Center for the Advancement of Topological Semimetals (CATS), an Energy Frontier Research Center funded by the USDOE Office of Science, Office of Basic Energy Sciences, through the Ames Laboratory.  Ames Laboratory is operated for the USDOE by Iowa State University under Contract No. DE-AC02-07CH11358. A portion of this research used resources at the Spallation Neutron Source, which is a USDOE Office of Science User Facility operated by the Oak Ridge National Laboratory.

\clearpage
\newpage
\setcounter{equation}{0}
\setcounter{figure}{0}
\setcounter{table}{0}
\setcounter{page}{7}
\setcounter{section}{0}

\renewcommand{\figurename}{Supplementary Figure}
\renewcommand{\tablename}{Supplementary Table}
\renewcommand{\author}{}
\renewcommand{\affiliation}{}
\renewcommand{\bibnumfmt}[1]{[S#1]}
\renewcommand{\citenumfont}[1]{S#1}
\renewcommand{\theequation}{S\arabic{equation}}

\begin{titlepage}
\begin{center}
\large{\textbf{{Chiral and flat-band magnetic cluster excitations in a ferromagnetic kagome metal \\ \normalfont (Supplementary Information)}}}

\bigbreak

\author{S. X. M. Riberolles}
\affiliation{Ames Laboratory, Ames, IA, 50011, USA}

\author{Tyler J. Slade}
\affiliation{Ames Laboratory, Ames, IA, 50011, USA}

\author{Bing Li}
\affiliation{Ames Laboratory, Ames, IA, 50011, USA}

\author {D.~L.~Abernathy}
\affiliation{Neutron Scattering Division, Oak Ridge National Laboratory, Oak Ridge, TN 37831 USA}

\author{P.~C.~Canfield}
\affiliation{Ames Laboratory, Ames, IA, 50011, USA}
\affiliation{Department of Physics and Astronomy, Iowa State University, Ames, IA, 50011, USA}

\author{B. G. Ueland}
\affiliation{Ames Laboratory, Ames, IA, 50011, USA}

\author{P. P. Orth}
\affiliation{Ames Laboratory, Ames, IA, 50011, USA}
\affiliation{Department of Physics and Astronomy, Iowa State University, Ames, IA, 50011, USA}

\author{Liqin Ke}
\affiliation{Ames Laboratory, Ames, IA, 50011, USA}

\author{R.~J.~McQueeney}
\affiliation{Ames Laboratory, Ames, IA, 50011, USA}
\affiliation{Department of Physics and Astronomy, Iowa State University, Ames, IA, 50011, USA}

 \date{\today}
\end{center} 
 \end{titlepage}
\
\bigbreak

 \section{Experimental Details}
 TbMn$_6$Sn$_6$ crystallizes in the HfFe$_6$Ge$_6$-type structure with hexagonal space group P6/$mmm$ (No.~191) and lattice parameters $a$ and $c$ are 5.530 and 9.023~\AA~ at 300 K \cite{Malaman99}. 
 
 \section{Heisenberg Model Description}
 
The Heisenberg model is given by  $\mathcal{H=H_{\rm intra}+H_{\rm inter}+H_{\rm aniso}}$.  Each Mn kagome layer possesses strong NN FM exchange ($\mathcal{J}_1=-28.8$ meV),
\begin{equation}
\mathcal{H}_{\rm intra}=\mathcal{J}_1\sum_{\langle i<j \rangle} \bold{s}_i \cdot \bold{s}_j
\label{Hintra}
\end{equation}
where $\bold{s}$ is the Mn spin operator with magnitude $s=1$. Several unique interlayer magnetic couplings between Mn layers and between Mn and Tb layers are found, giving
\begin{equation}
\mathcal{H}_{\rm inter}=\sum_{k} \sum_{i<j} \mathcal{J}^{MM}_k\bold{s}_i \cdot \bold{s}_{j+k} + \mathcal{J}^{MT} \sum_{\langle i<j \rangle} \bold{s}_i \cdot \bold{S}_j.
\label{Hinter}
\end{equation}
Here, $\mathcal{J}^{MT}=0.93$ meV is the AF coupling between neighboring Mn and Tb layers, with Tb having a total spin angular momentum of $S=3$ and Mn having $s=1$.  Competing interactions between Mn layers with a layer index $k$ up to $k=3$ ($\mathcal{J}^{MM}_1= -4.4$ meV, $\mathcal{J}^{MM}_2=-19.2$ meV, $\mathcal{J}^{MM}_3= 1.8$ meV) are necessary to fully describe the low energy spin waves, especially the splitting of AO and AE modes. Finally, uniaxial and easy-plane single-ion anisotropies for Tb and Mn, respectively, are described by
\begin{equation}
\mathcal{H}_{\rm aniso}= K^T \sum_{i} (S_i^z)^2 + K^M \sum_{i} (s_i^z)^2
\label{Haniso}
\end{equation}
 with $K^T=-1.28$ meV and $K^M=+0.44$ meV \cite{Riberolles22}.

Extended models where longer-range intralayer Mn-Mn interactions have been introduced, such as $J_2$, $J_{3a}$ (through the Mn atom), and $J_{3b}$ (across the hexagon).  Generally, longer-range FM interactions pull down the average energy OE and OO modes in better correspondence to experiments. We try not to affect the AO and AE modes by introducing constraints. For example, keeping (such as $J_1+3J_2=constant$) will keep the M-point energies of the AE and AO modes fixed.  Fig.~\ref{dispersion} shows some results for the extended models.
 \begin{figure}
\includegraphics[width=1.0\linewidth]{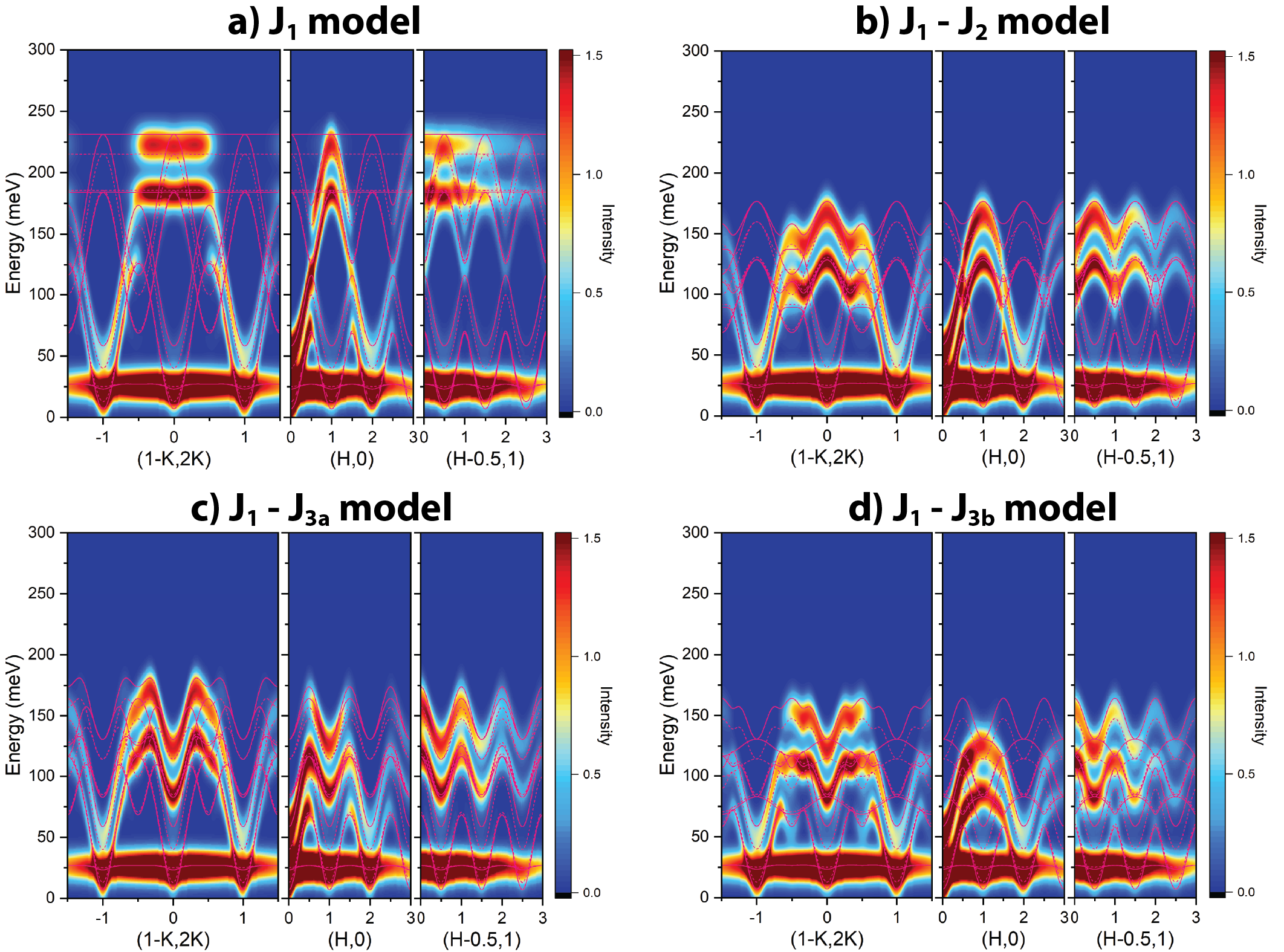}
\caption{\footnotesize Calculations of the neutron scattering intensity along the (1-K,2K), (H,0) and (H-0.5,1) directions for Heisenberg model with (a) Nearest-neighbor intralayer interaction $J_1= -28.8$ meV, (b) nearest and next-nearest neighbor interactions with $J_1=-15$ meV and $J_2=-4.6$ meV and $J_1+3J_2=-28.8$ meV, (c) nearest and third-nearest neighbor (through the Mn) interactions with $J_1=-12$ meV and $J_{3a}=-8.4$ meV and $J_1+2J_{3a}=-28.8$ meV, (d) nearest and third-nearest neighbor (across the hexagon) interactions with $J_1=-12$ meV and $J_{3b}=-8.4$ meV and $J_1+2J_{3b}=-28.8$ meV. }
\label{dispersion}
\end{figure} 

In Fig. \ref{constE}, we show constant energy slices from the Heisenberg model parameters used in the main text.  This can be compared to the data shown in Fig. 2 of the main text.  The model calculations indicate that intensity principally occurs in the $\Gamma$ zones below 130 meV (comprised of AE and AO modes) and only in the $\Gamma'$ zones above 130 meV (comprised of OE,OO,FE, and FO modes). The flat band modes that occur at the highest energies have intensities centered at the K-point.

 \begin{figure*}
\includegraphics[width=1.0\linewidth]{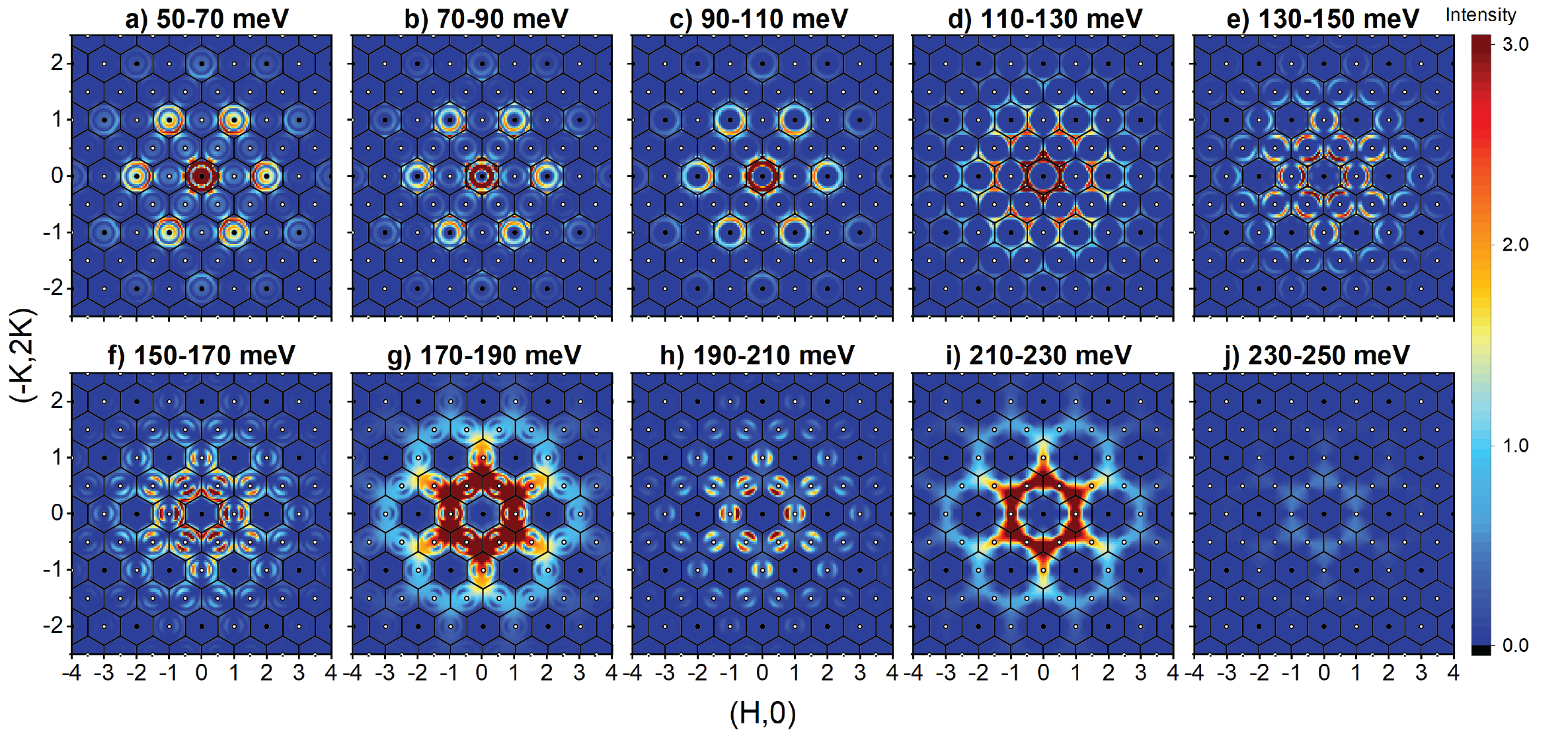}
\caption{\footnotesize Constant energy slices the TbMn$_6$Sn$_6$ spin excitations from a Heisenberg model. Slices are identical to the data cuts shown in Fig. 2 and are made over incremental energy ranges from (a) 50--70 meV, (b) 70--90 meV, (c) 90--110 meV, (d) 110--130 meV, (e) 130--150 meV, (f) 150--170 meV, (g) 170--190 meV, (h) 190--210 meV, (i) 210--230 meV, (j) 230--250 meV.  Data were collected at $E_i=250$ meV for panels (a)--(e) and $E_i=500$ meV for panels (f)--(j). All cuts are averaged over an $L$-range from -7 to 7 rlu. Hexagonal Brillouin zone boundaries are shown and $\Gamma$ and $\Gamma'$ zone centers are indicated by filled and empty circles, respectively.}
\label{constE}
\end{figure*} 

\section{Damped harmonic oscillator analysis}
\begin{figure}
\includegraphics[width=0.75\linewidth]{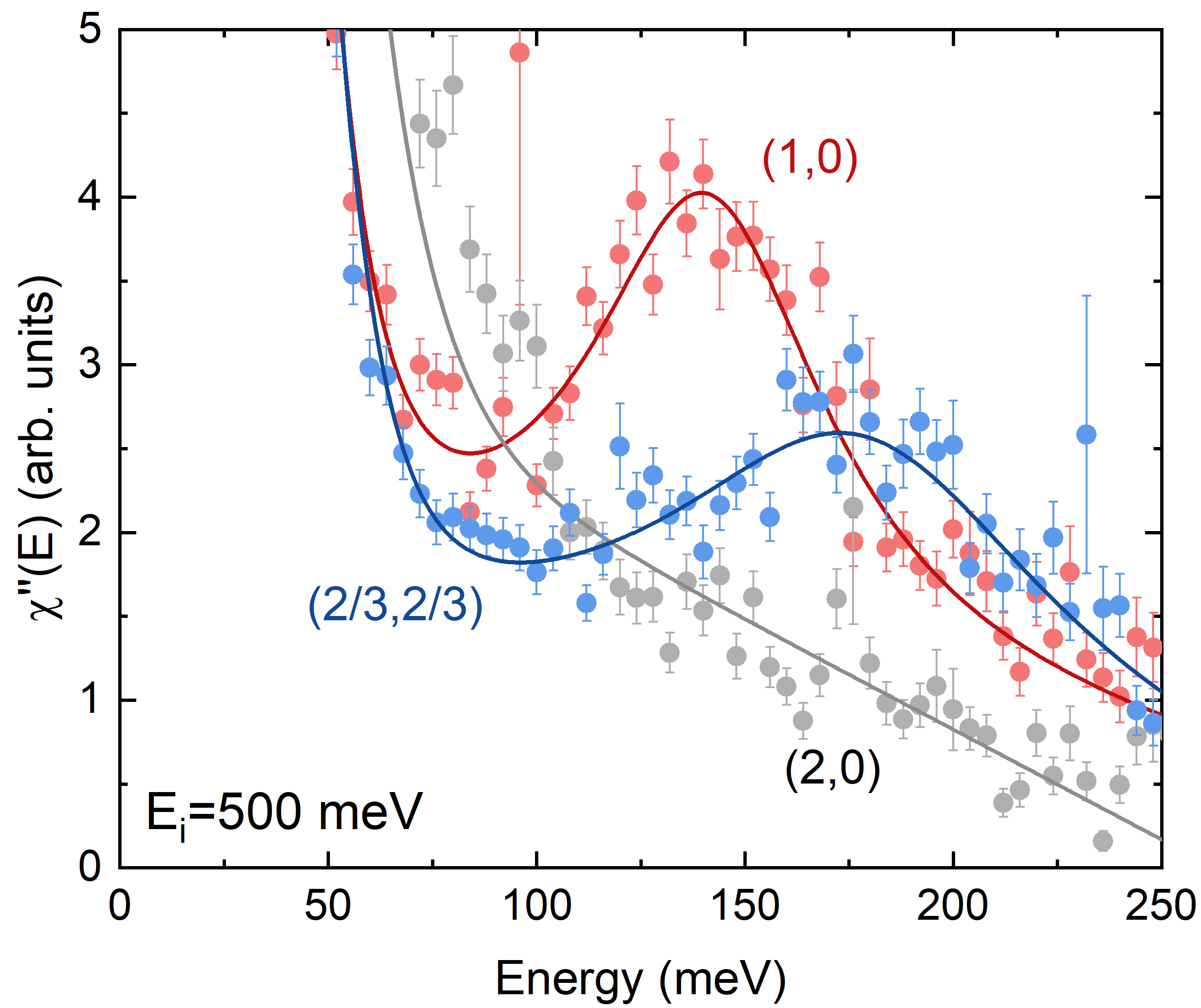}
\caption{\footnotesize Energy cuts centered at the $\Gamma'=(1,0)$ point (red), at the ${\rm K}=(\frac{2}{3},\frac{2}{3})$ point (blue), and at the $\Gamma=(2,0)$ point (gray) with $E_i=$ 500 meV.  The data are scaled to the imaginary part of the dynamical susceptibility, $\chi''(E)$, after correcting for the $L$-averaged magnetic form factor.  Solid lines correspond to damped harmonic oscillator fits, as described in the text.}
\label{cuts}
\end{figure}

Figure \ref{cuts} shows cuts at the key reciprocal space positions for the chiral and flat band excitations. The dynamical susceptibility is fit to a damped harmonic oscillator lineshape, $\chi''(E)=A\gamma E/[(E^2-\omega_i^2)^2+E^2\gamma_i^2]$, plus a background.  We find the mean frequencies of the chiral and flat mode cluster excitations, $\omega_C=$145(2) meV and $\omega_F=$187(8) meV, and the corresponding relaxation rates, $\gamma_C=$69(10) meV and $\gamma_F=$115(30) meV, respectively.  The relaxation rates are influenced by the background fitting, so we used the cut at $\Gamma=(2,0)$ to estimate of the high-energy background level. Nonetheless, an oscillator $Q$-factor of $\omega/\gamma\approx1.5-2$ for the two modes is consistent with large damping.

 \section{Temperature dependence}
\begin{figure*}
\includegraphics[width=1.0\linewidth]{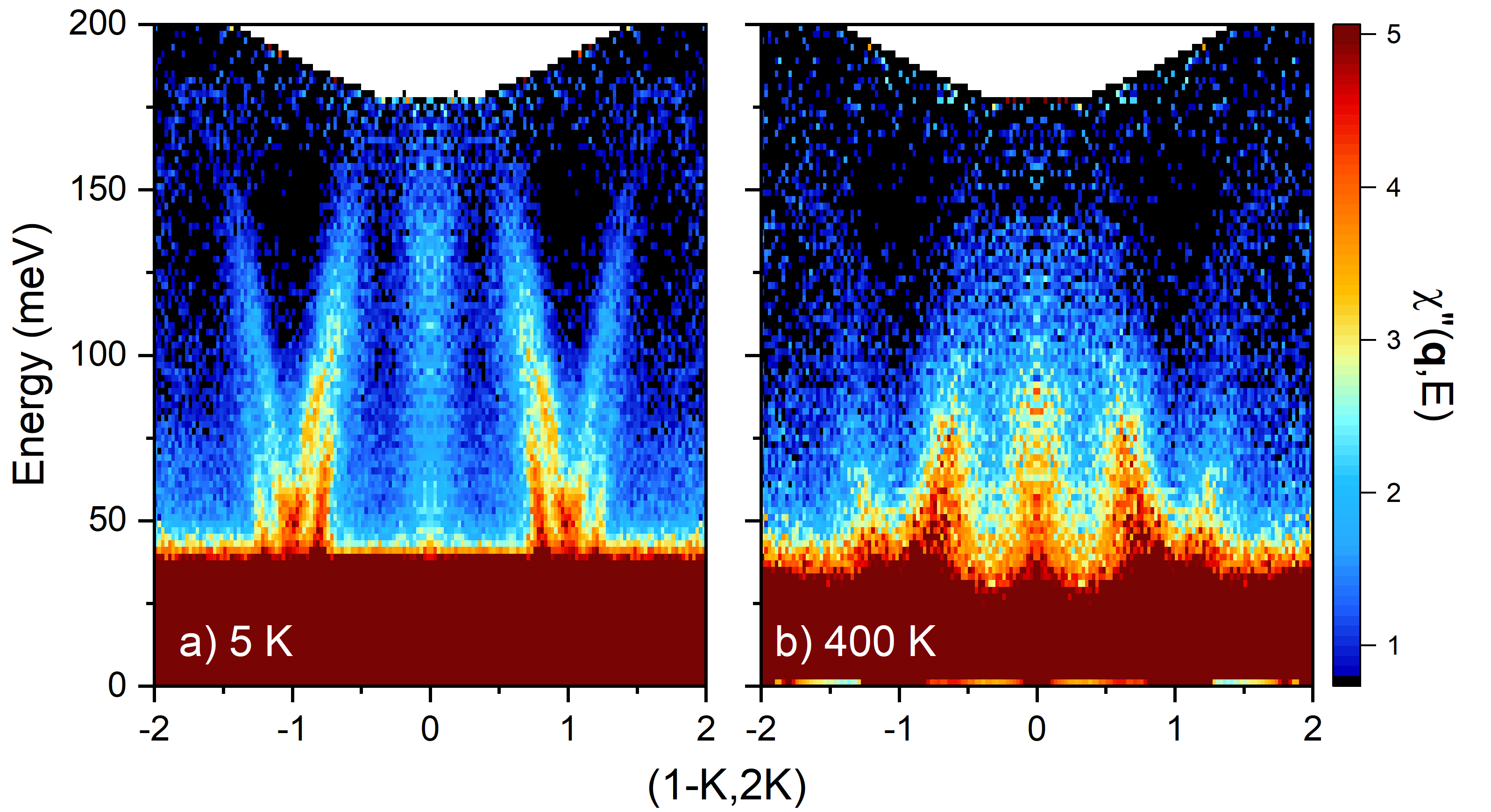}
\caption{\footnotesize Comparison of slices along the $(-K,2K,0)$ direction for a) $T=5$ K and b) 400 K.  Data are scaled to be proportional to the dynamical susceptibility, $\chi''({\bf q},E)=I({\bf q},E)(1-{\rm exp}(-E/kT))$.  Cuts are performed with $E_i=250$ meV.  All plots are also averaged over reciprocal space ranges of $H=[0.9:1.1]$ and $L=[-7:7]$.  Data are symmetrized according to the $P6/mmm$ space group to improve statistics. }
\label{slice_temperature}
\end{figure*}

 Figure \ref{slice_temperature} compares slices through excitations in the $\Gamma$ and $\Gamma'$ zones at $T=5$ K and 400 K as measured with $E_i=250$ meV for several energy transfers.  The data show that the dispersive AO and AE excitations in the (-1,2,0) $\Gamma$ zone are further split at 400 K, indicating significant softening of the Mn modes just below $T_C$.  The localized $q=0$ cluster mode in the (1,0,0) $\Gamma'$ zone is still observed at 400 K. Both the dispersive $\Gamma$ and localized $\Gamma'$ modes are more intense at lower energies, but their intensity rapidly diminishes above 100 meV as a consequence of softening and mode damping near $T_C$.  Evidence of the dispersive $\Gamma$ mode is present at 130 meV.  However, the $\Gamma'$ excitation appears to be completely suppressed above 130 meV at 400 K.  Despite the complex temperature dependence, which also must take into account the transition from uniaxial to easy-plane ferrimagnetism above $T_{SR}=310$ K, the data suggest that the both the $\Gamma$ and $\Gamma'$ excitations occur on an ordered Mn sublattice.

\section{Cluster calculations}
\begin{figure}
\includegraphics[width=0.5\linewidth]{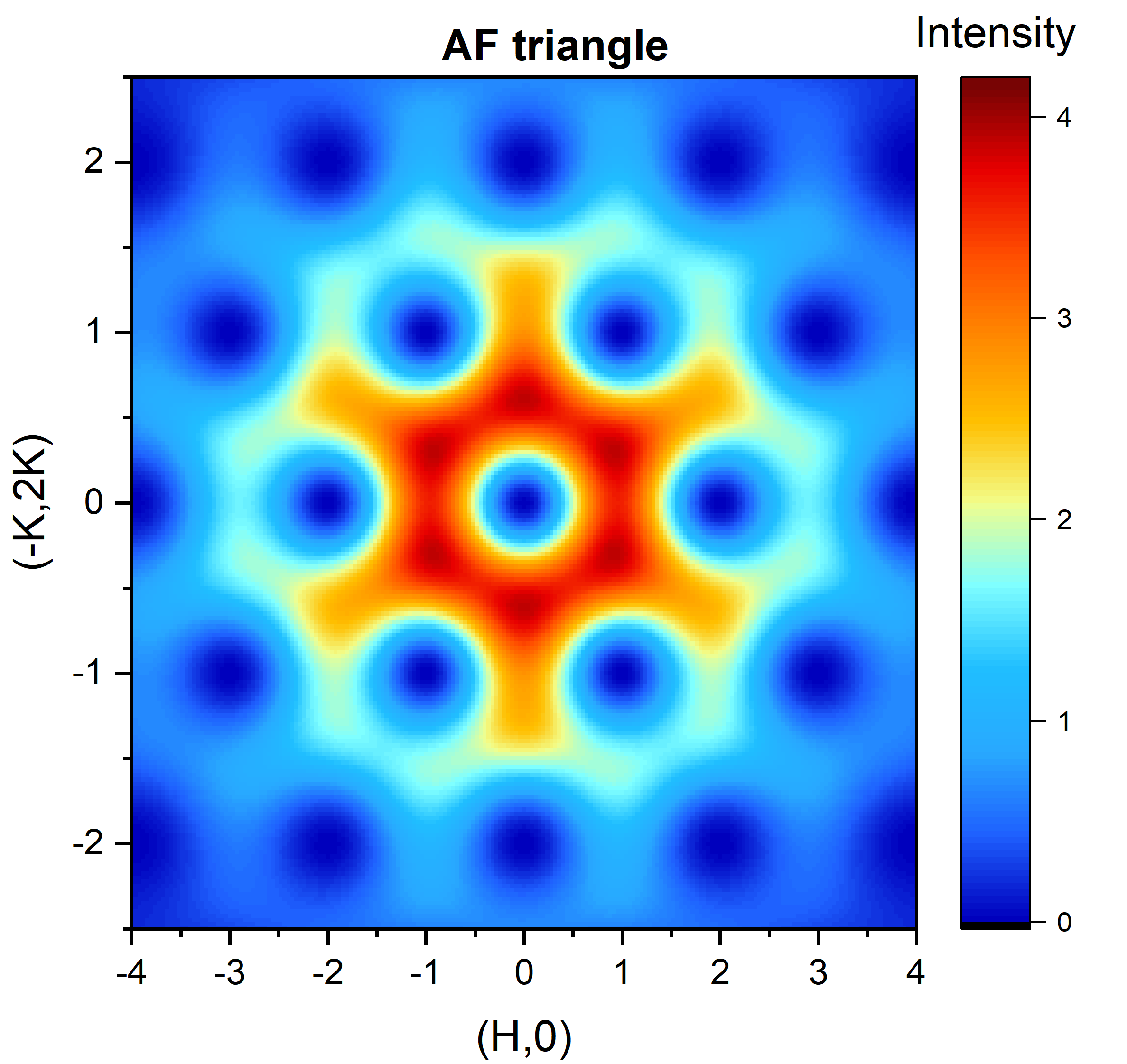}
\caption{\footnotesize Structure factor of a triangular cluster with chiral $\Gamma'$ spin correlations.}
\label{fig:triangle}
\end{figure}

\begin{figure}
\includegraphics[width=1.0\linewidth]{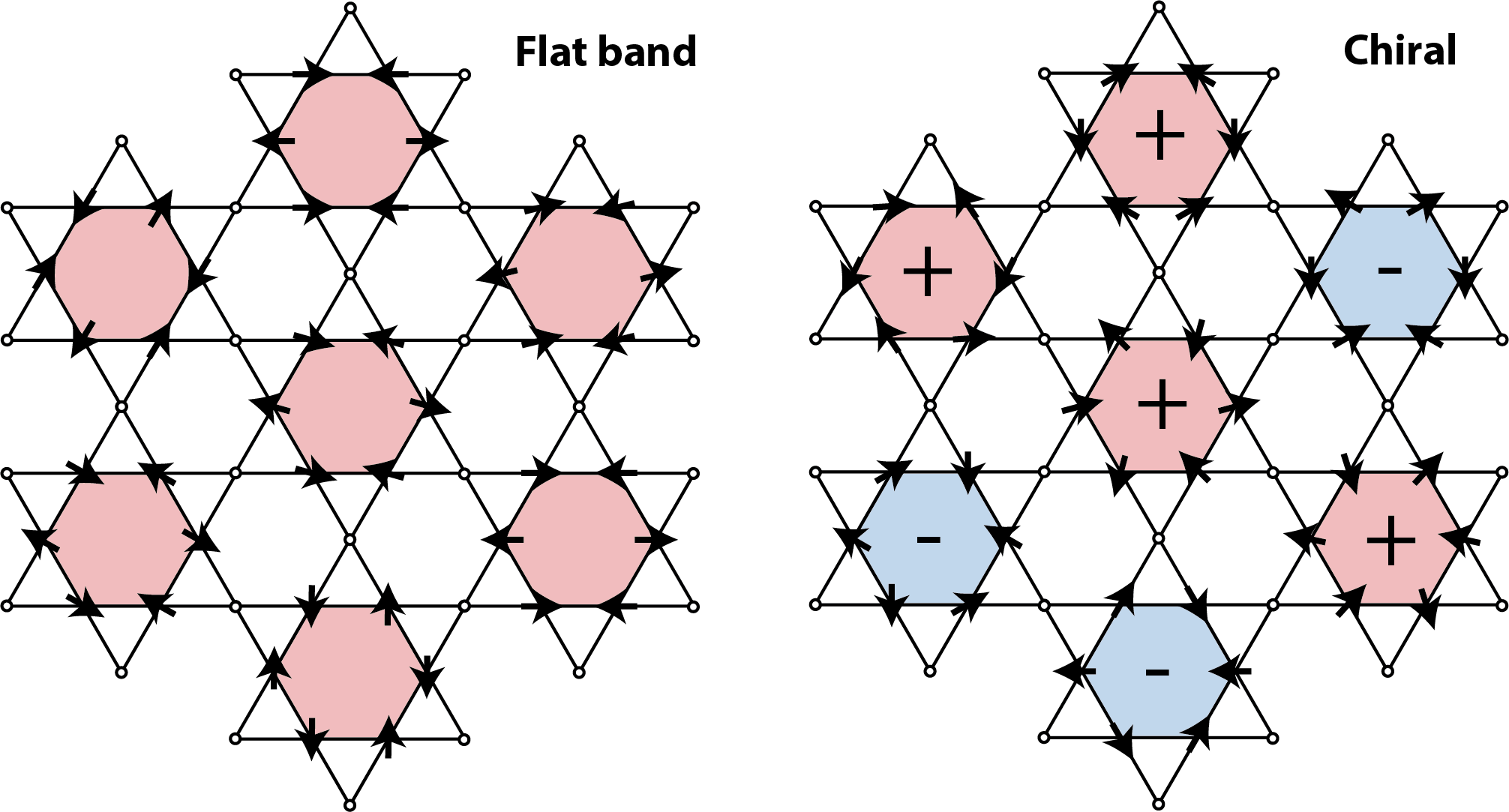}
\caption{\footnotesize Extended picture of flat band and chiral excitations on hexagonal clusters.  The chiral excitations have a handedness indicated by the color of the plaquette and the sign. }
\label{fig:cartoon}
\end{figure}

Spin correlations are most likely to occur in hexagonal or trangular plaquettes in the kagome layer.  In Fig.~\ref{fig:triangle}, we plot the structure factor for chiral spin correlations around a triangular plaquette.  This pattern is not consistent with the $q$-dependence of the data in the $\Gamma'$ zones shown in Fig. 2(a)--(e).  Fig.~\ref{fig:cartoon} shows a schematic diagram of flat band and chiral spin excitations which have no phase coherence from hexagon-to-hexagon.

\section{Density-functional theory calculations and Landau damping}

\begin{figure}[ht]
\centering
\begin{tabular}{cc}
  \includegraphics[width=.40\linewidth,clip]{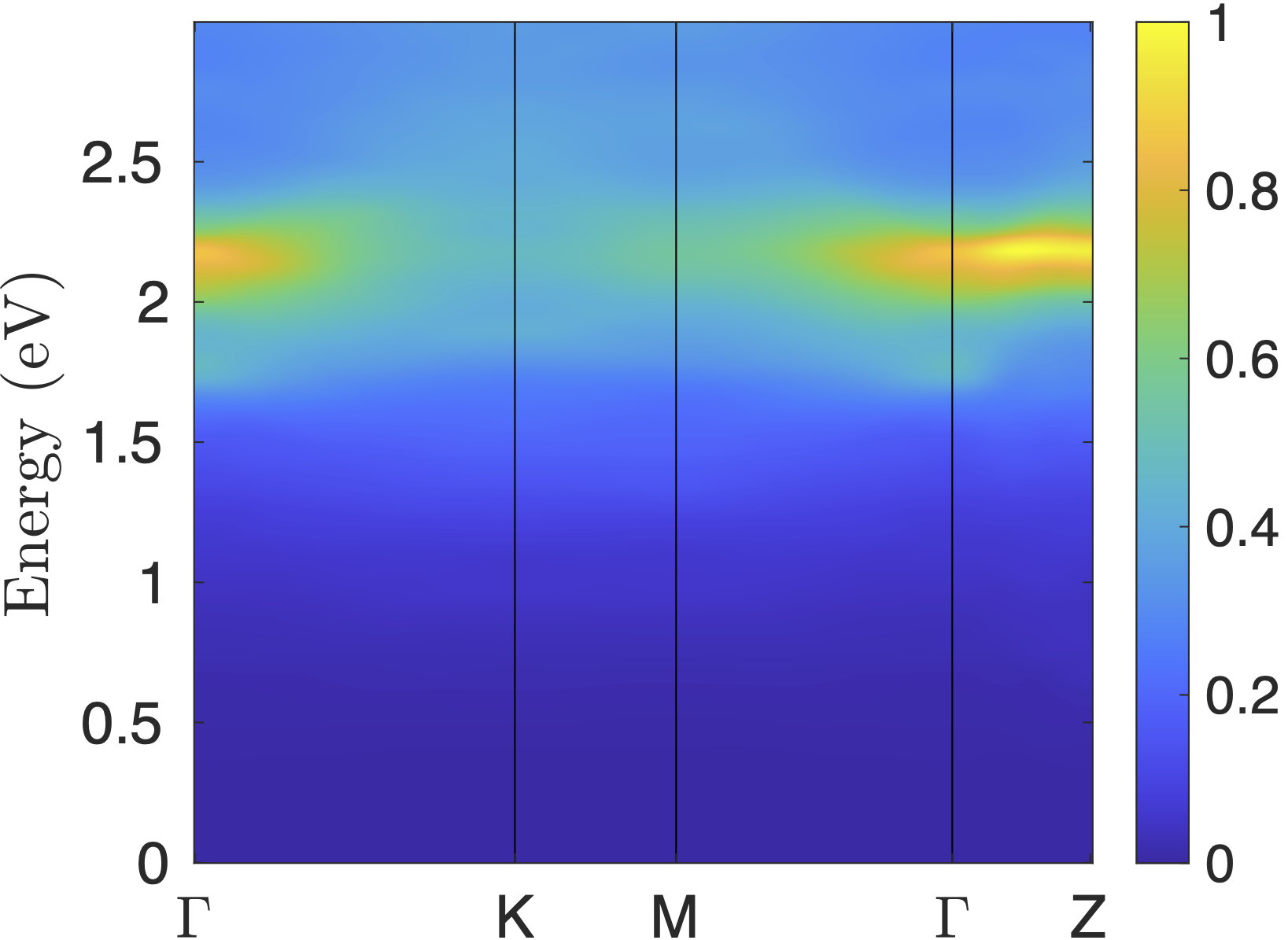} &
  \includegraphics[width=.42\linewidth,clip]{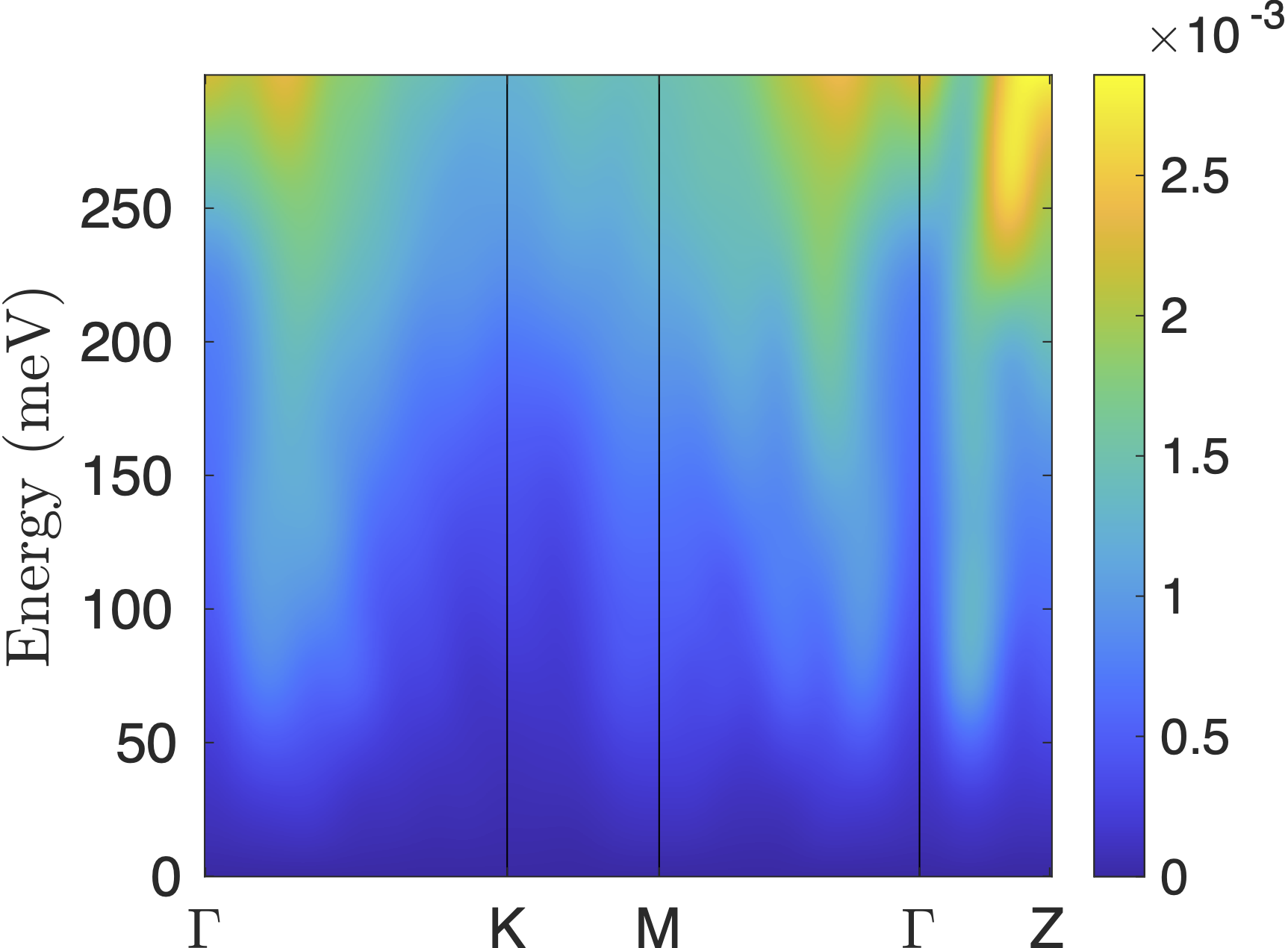}   
\end{tabular}%
\caption{The imaginary part of bare transverse spin susceptibility $\operatorname{Im}[\xqwtb]$ along the high symmetry path $\Gamma$--K--M--$\Gamma$--Z in TbMn$_6$Sn$_6$. 
The intensity of $\operatorname{Im}[\xqwtb]$, normalized with $\text{max}(\operatorname{Im}[\xqwtb])$, is shown in the energy windows of 0--3~eV (left panel) and 0--300~meV (right panel), respectively.
$\operatorname{Im}[\xqwtb]$ characterizes the intensity of single-particle spin-flip excitations---the Stoner excitations.}
\label{fig:x0}
\end{figure}

\textit{Band structure}.
The electronic structures of TbMn$_6$Sn$_6$ are calculated in DFT+$U$ using an all-electron full-potential LMTO (FP-LMTO) \cite{Methfessel00}.
Plain DFT usually incorrectly positions the Lanthanide 4f states near the Fermi level.
A sizable on-site Hubbard $U\approx7$~eV correction was applied on Tb-$f$ states to push down the occupied $4f$ states and further away from the Fermi level.
The resulting spin magnetic moments and bandstructure compare well with those previously calculated using other full-potential all-electron methods~\cite{lee2022arxiv}.

\textit{Susceptibility calculation}.
After obtaining the self-consistent \textit{ab initio} Hamiltonian $H$, we calculate the bare transverse spin susceptibility $\xqwtb$ using the eigenvalues and eigenfunctions of $H$ within a linear response theory~\cite{kotani2008jpcm,ke2011prbr,ke2021ncm,li2021prb},

\begin{eqnarray}
\begin{aligned}
\chi_0^{+-}(\bfr,\bfr',\bfq, \omega) &=& \\
 \sum^{\rm  occ}_{\bfk n \ispone} \sum^{\rm unocc}_{\bfk' n'\isptwo}
\frac{
\Psi_{\bfk n\ispone}^*(\bfr)      \Psi_{\bfk' n'\isptwo}(\bfr)
\Psi_{\bfk' n'\isptwo}^*(\bfr') \Psi_{\bfk n\ispone}(\bfr')
}{\omega-(\epsilon_{\bfk' n'\isptwo}-\epsilon_{\bfk n\ispone})+i \delta} \nonumber\\ 
&+& \\
\sum^{\rm  unocc}_{\bfk n \ispone} \sum^{\rm occ}_{\bfk' n'\isptwo}
\frac{
\Psi_{\bfk n\ispone}^*(\bfr)      \Psi_{\bfk' n'\isptwo}(\bfr)
\Psi_{\bfk' n'\isptwo}^*(\bfr') \Psi_{\bfk n\ispone}(\bfr')
}{-\omega-(\epsilon_{\bfk n\ispone}-\epsilon_{\bfk' n'\isptwo})+i \delta},~~~(S4)
\end{aligned}
\label{generalchi01q}
\end{eqnarray}

where $\bfk' = \bfq+ \bfk$.
For computational efficiency, $\chi_0$ are calculated using a mixed basis~\cite{kotani2007prb}, which consists of the product basis~\cite{aryasetiawan1994prbA,kotani2007prb} within the augmentation spheres and interstitial plane waves. 
We first calculate $\chi_0^{+-}(\bfq,\omega)$ on a $12\times12\times6$ $\bfq$ mesh and construct the $\chi_0^{+-}(\bfR,\omega)$ through Fourier transformation.
Then we use the resulted $\chi_0^{+-}(\bfR,\omega)$ to calculate $\chi_0^{+-}(\bfq,\omega)$ with a dense set of $\bfq$ points along high-symmetry paths.

The Mn-$d$ bands dominate near the Fermi level, governing the spin-flip excitations.
Figure ~\ref{fig:x0} shows the imaginary part of bare transverse spin susceptibility $\operatorname{Im}[\xqwtb]$ along the high symmetry path $\Gamma$--K--M--$\Gamma$--Z.
As shown in the left panel, the most prominent Stoner excitations occur near 2.2~eV, corresponding to the exchange splitting of Mn-$3d$ bands that dominate near the Fermi level.
However, since TbMn$_6$Sn$_6$ is a metallic system, small finite ${\bf q}$-dependent $\xqwtb$ can also be observed in the energy window of spinwave excitations (right panel), which may cause the spinwave damping.

\begin{figure}
\includegraphics[width=0.8\linewidth]{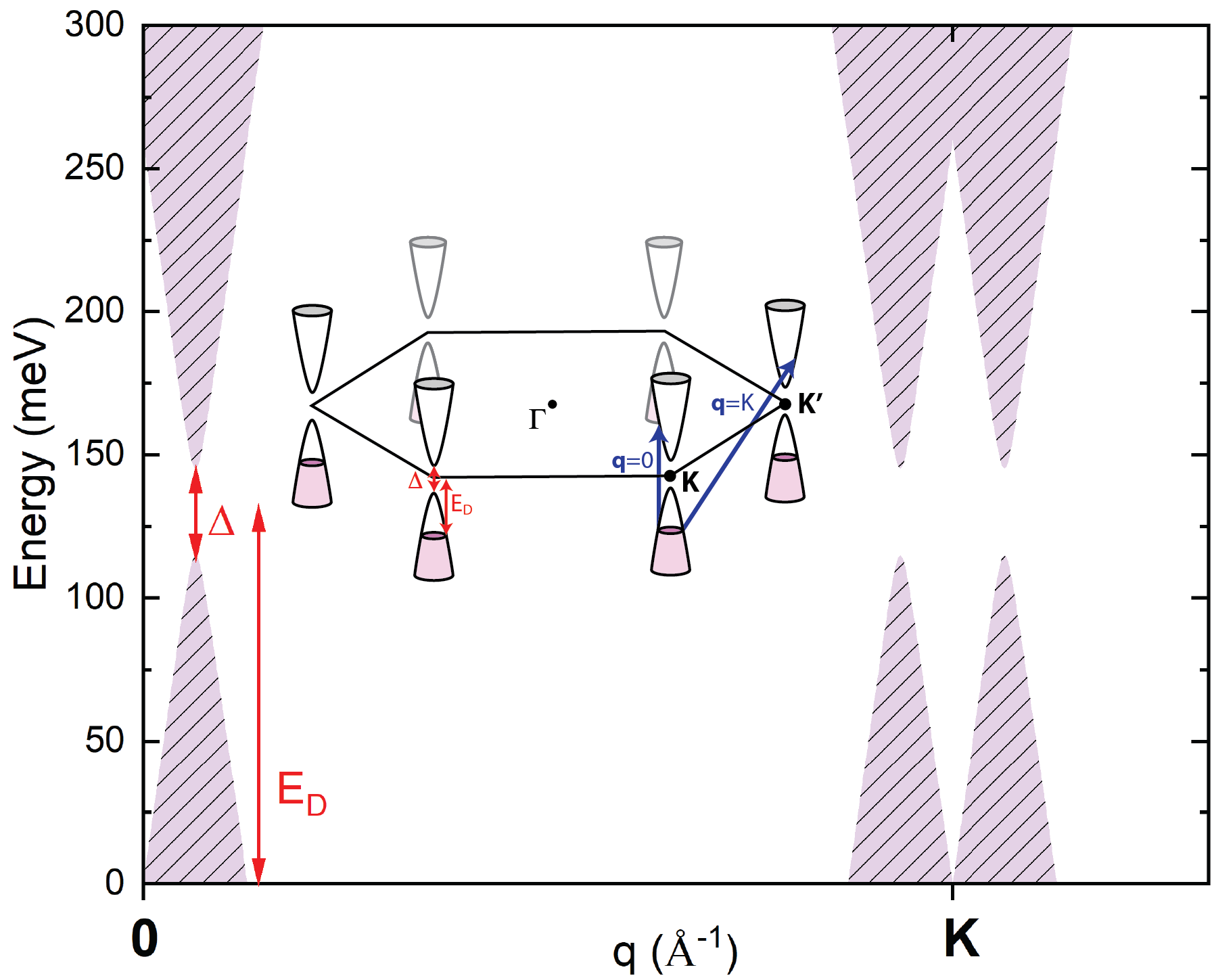}
\caption{\footnotesize Continuum of electron-hole scattering channels for massive Dirac fermions in a 2D kagome layer. Dirac fermions are characterized by the chemical potential $E_D$ and gap $\Delta$.}
\label{fig:mdf}
\end{figure}

\textit{Massive Dirac fermions}.
Figure \ref{fig:mdf}, shows the expectation for intravalley and intervalley charge scattering for massive Dirac fermions (MDFs), which form a continuum of scattering near the $\Gamma$ and K-points.  Experimental data place the chemical potential at $E_D=-130$ meV with a band gap of $\Delta\approx30$ meV \cite{Ma21}.  The features of the continuum of scattering from interband transitions of the MDFs show similar energy onset and $q$-dependence as the data.  However, the spin-polarized nature of the MDFs indicates that these are not spin-flip transitions and therefore should not couple to the magnon.  This assumption is supported by our DFT calculations.

\section{Acknowledgments} RJM, LK, PPO, BGU, BL, and SXMR's work at the Ames Laboratory is supported by the U.S. Department of Energy (USDOE), Office of Basic Energy Sciences, Division of Materials Sciences and Engineering. TJS, TH, and PC are supported by the Center for the Advancement of Topological Semimetals (CATS), an Energy Frontier Research Center funded by the USDOE Office of Science, Office of Basic Energy Sciences, through the Ames Laboratory.  Ames Laboratory is operated for the USDOE by Iowa State University under Contract No. DE-AC02-07CH11358. A portion of this research used resources at the Spallation Neutron Source, which is a USDOE Office of Science User Facility operated by the Oak Ridge National Laboratory.


\begin{thebibliography}{99}
\bibitem{Ortiz20} B. R. Ortiz, S. M. L. Teicher, Y. Hu, J. L. Zuo, P. M. Sarte, E. C. Schueller, A. M. M. Abeykoon, M. J. Krogstad, S. Rosenkranz, R. Osborn et al., "$\mathrm{Cs}{\mathrm{V}}_{3}{\mathrm{Sb}}_{5}$: A ${\mathbb{Z}}_{2}$ Topological Kagome Metal with a Superconducting Ground State", Phys. Rev. Lett. {\bf 125}, 247002 (2020).
\bibitem{Jiang21} Y.-X. Jiang, J.-X. Yin, M. M. Denner, N. Shumiya, B. R. Ortiz, G. Xu, Z. Guguchia, J. He, M. S. Hossain, X. Liu et al., "Unconventional chiral charge order in kagome superconductor KV$_3$Sb$_5$", Nat. Mater. {\bf 20}, 1353 (2021).
\bibitem{Liu18} E. Liu, Y. Sun, N. Kumar, L. Muechler, A. Sun, L. Jiao, S.-Y. Yang, D. Liu, A. Liang, Q. Xu et al., "Giant anomalous Hall effect in a ferromagnetic kagome-lattice semimetal", Nat. Phys. {\bf 14}, 1125 (2018).
\bibitem{Yin22} J.-X. Yin, B. Lian and M. Z. Hasan, "Topological kagome magnets and superconductors" Nature {\bf 612}, 647 (2022).

\bibitem{Ghimire20} N. J. Ghimire, R. L. Dally, L. Poudel, D. C. Jones, D. Michel, N. T. Magar, M. Bleuel, M. A. McGuire, J. S. Jiang, J. F. Mitchell et al., "Competing magnetic phases and fluctuation-driven scalar spin chirality in the kagome metal YMn$_6$Sn$_6$", Sci. Adv. {\bf 6}, eabe2680 (2020).
\bibitem{Yin20} J.-X. Yin, W. Ma, T. A. Cochran, X. Xu, S. S. Zhang, H.-J. Tien, N. Shumiya, G. Cheng, K. Jiang, B. Lian et al., "Quantum-limit Chern topological magnetism in TbMn$_6$Sn$_6$", Nature {\bf 583}, 533 (2020).
\bibitem{Ma21} W. Ma, X. Xu, J.-X. Yin, H. Yang, H. Zhou, Z.-J. Cheng, Y. Huang, Z. Qu, F. Wang, M. Z. Hasan et al., ''Rare Earth Engineering in $R$Mn$_6$Sn$_6$ ($R=$Gd-Tm,Lu) Topological Kagome Magnets'', Phys. Rev. Lett. {\bf 126}, 246602 (2021).
\bibitem{Li21}M. Li, Q. Wang, G. Wang, Z. Yuan, W. Song, R. Lou, Z. Liu, Y. Huang, Z. Liu, H. Lei et al., ''Dirac cone, flat band and saddle point in kagome magnet YMn$_6$Sn$_6$'', Nat. Commun. {\bf 12}, 3129 (2021).
\bibitem{Dhakal21} G. Dhakal, F. Cheenicode Kabeer, A. K. Pathak, F. Kabir, N. Poudel, R. Filippone, J. Casey, A. Pradhan Sakhya, S. Regmi, C. Sims et al., ''Anisotropically large anomalous and topological Hall effect in a kagome magnet'', Phys. Rev. B {\bf 104}, L161115 (2021).
\bibitem{Kabir22} F. Kabir, R. Filippone, G. Dhakal, Y. Lee, N. Poudel, J. Casey, A. P. Sakhya, S. Regmi, R. Smith, P. Manfrinetti et al., ''Unusual magnetic and transport properties in HoMn$_6$Sn$_6$ kagome magnet'', Phys. Rev. Mater. {\bf 6}, 064404 (2022).
\bibitem{Idrissi91} B. C. El Idrissi, G. Venturini, B. Malaman, and D. Fruchart, "Magnetic structures of TbMn$_6$Sn$_6$ and HoMn$_6$Sn$_6$ compounds from neutron diffraction study", J.   Less-Common Met. {\bf 175}, 143 (1991).

\bibitem{Kakihana18} M. Kakihana, K. Nishimura, D. Aoki, A. Nakamura, M. Nakashima, Y. Amako, T. Takeuchi, T. Kida, T. Tahara, M. Hagiwara et al., ''Electronic States of Antiferromagnet FeSn and Pauli Paramagnet CoSn'', J. Phys. Soc. Jpn. {\bf 88}, 014705 (2018).
\bibitem{Kang20} M. Kang, L. Ye, S. Fang, J.-S. You, A. Levitan, M. Han, J. I. Facio, C. Jozwiak, A. Bostwick, E. Rotenberg et al., ''Dirac fermions and flat bands in the ideal kagome metal FeSn'', Nat. Mater. {\bf 19}, 163 (2020).
\bibitem{Ye20} L. Ye, M. Kang, J. Liu, F. von Cube, C. R. Wicker, T. Suzuki, C. Jozwiak, A. Bostwick, E. Rotenberg, D. C. Bell et al., ''Massive Dirac fermions in a ferromagnetic kagome metal'', Nature {\bf 555}, 638 (2018).
\bibitem{Han21} M. Han, H. Inoue, S. Fang, C. John, L. Ye, M. K. Chan, D. Graf, T. Suzuki, M. P. Ghimire, W. J. Cho et al., ''Evidence of two-dimensional flat band at the surface of antiferromagnetic kagome metal FeSn'', Nat. Commun. {\bf 12}, 5345 (2021).
\bibitem{Kida11} T. Kida, L. A. Fenner, A. A. Dee, I. Terasaki, M. Hagiwara, and A. S. Wills, ''The giant anomalous Hall effect in the ferromagnet Fe$_3$Sn$_2$ a frustrated kagome metal'', J. Phys.:Condens. Matter {\bf 23}, 112205 (2011).
\bibitem{Lin18} Z. Lin, J.-H. Choi, Q. Zhang, W. Qin, S. Yi, P. Wang, L. Li, Y. Wang, H. Zhang, Z. Sun et al., ''Flatbands and Emergent Ferromagnetic Ordering in Fe$_3$Sn$_2$ Kagome Lattices'', Phys. Rev. Lett. {\bf 121}, 096401 (2018).
\bibitem{Ye19} L. Ye, M. K. Chan, R. D. McDonald, D. Graf, M. Kang, J. Liu, T. Suzuki, R. Comin, L. Fu, and J. G. Checkelsky, ''de Haas-van Alphen effect of correlated Dirac states in kagome metal Fe$_3$Sn$_2$'', Nat. Commun. {\bf 10}, 4870 (2019).

\bibitem{Mook14}  A. Mook, J. Henk, and I. Mertig, ''Magnon Hall effect and topology in kagome lattices: A theoretical investigation'', Phys. Rev. B {\bf 89}, 134409 (2014).
\bibitem{Onose10}Y. Onose, T. Ideue, H. Katsura, Y. Shiomi, N. Nagaosa, and Y. Tokura, ''Observation of the Magnon Hall Effect'', Science {\bf 329}, 297 (2010).
\bibitem{Mook14_2} A. Mook, J. Henk, and I. Mertig, ''Edge states in topological magnon insulators'', Phys. Rev. B {\bf 90}, 024412 (2014).

\bibitem{Zhang20} H. Zhang, X. Feng, T. Heitmann, A. I. Kolesnikov, M. B. Stone, Y. M. Lu, and X. Ke, "Topological magnon bands in a room-temperature kagome magnet", Phys. Rev. B {\bf 101}, 100405 (2020).
\bibitem{Xie21} Y. Xie, L. Chen, T. Chen, Q. Wang, Q. Yin, J. R. Stewart, M. B. Stone, L. L. Daemen, E. Feng, H. Cao {\it et al.}, "Spin excitations in metallic kagome lattice FeSn and CoSn", Commun. Phys {\bf 4}, 240 (2021).
\bibitem{Do21} S.-H. Do, K. Kaneko, R. Kajimoto, K. Kamazawa, M. B. Stone, J. Y. Y. Lin, S. Itoh, T. Masuda, G. D. Samolyuk, E. Dagotto et al., ''Damped Dirac magnon in the metallic kagome antiferromagnet FeSn'', Phys. Rev. B {\bf 105}, L180403 (2022).
\bibitem{Riberolles22}  S. X. M. Riberolles, Tyler J. Slade, D. L. Abernathy, G. E. Granroth, Bing Li, Y. Lee, P. C. Canfield, B. G. Ueland, Liqin Ke, R. J. McQueeney, "Low-Temperature Competing Magnetic Energy Scales in the Topological Ferrimagnet TbMn$_6$Sn$_6$", Phys. Rev. X, {\bf 12}, 021043 (2022).

\bibitem{Bergman08} D. L. Bergman, C. Wu, and L. Balents, "Band touching from real-space topology in frustrated hopping models", Phys. Rev. B {\bf 78}, 125104 (2008).

\bibitem{Inami00} T. Inami, M. Nishiyama, S. Maegawa, and Y. Oka, "Magnetic structure of the kagom\'e lattice antiferromagnet potassium jarosite ${\mathrm{KFe}}_{3}{(\mathrm{OH})}_{6}{(\mathrm{S}\mathrm{O}}_{4}{)}_{2}$", Phys. Rev. B {\bf 61}, 12181 (2000).
\bibitem{Grohol05} D. Grohol, K. Matan, J.-H. Cho, S.-H. Lee, J. W. Lynn, D. G. Nocera, and Y. S. Lee, "Spin chirality on a two-dimensional frustrated lattice", Nat. Mater. {\bf 4}, 323 (2005).

\bibitem{Fawcett88} E. Fawcett, ''Spin-density-wave antiferromagnetism in chromium'', Rev. Mod. Phys. {\bf 60}, 209 (1988).
\bibitem{Diallo09} S. O. Diallo, V. P. Antropov, T. G. Perring, C. Broholm, J. J. Pulikkotil, N. Ni, S. L. Bud'ko, P. C. Canfield, A. Kreyssig, A. I. Goldman et al., ''Itinerant Magnetic Excitations in Antiferromagnetic CaFe$_2$As$_2$'', Phys. Rev. Lett. {\bf 102}, 187206 (2009).
\bibitem{Chen20} X. Chen, I. Krivenko, M. B. Stone, A. I. Kolesnikov, T. Wolf, D. Reznik, K. S. Bedell, F. Lechermann, and S. D. Wilson, ''Unconventional Hund metal in a weak itinerant ferromagnet'', Nat. Commun. {\bf 11}, 3076 (2020).

\bibitem{Liu21} Z. Liu, N. Zhao, M. Li, Q. Yin, Q. Wang, Z. Liu, D. Shen, Y. Huang, H. Lei, K. Liu et al., ''Electronic correlation effects in the kagome magnet GdMn$_6$Mn$_6$'', Phys. Rev. B {\bf 104}, 115122 (2021).
\bibitem{Gu22} X. Gu, C. Chen, W. S. Wei, L. L. Gao, J. Y. Liu, X. Du, D. Pei, J. S. Zhou, R. Z. Xu, Z. X. Yin et al., ''Robust kagome electronic structure in the topological quantum magnets $X$Mn$_6$Sn$_6$ ($X=$ Dy, Tb, Gd, Y)'', Phys. Rev. B {\bf 105}, 155108 (2022).
\bibitem{Mielke22} C. Mielke, D. Das, J. X. Yin, H. Liu, R. Gupta, Y. X. Jiang, M. Medarde, X. Wu, H. C. Lei, J. Chang et al., ''Time-reversal symmetry-breaking charge order in a kagome superconductor", Nature {\bf 602}, 245 (2022).
\bibitem{SM} Supplementary Material link placeholder.

\bibitem{Chisnell15} R. Chisnell, J. S. Helton, D. E. Freedman, D. K. Singh, R. I. Bewley, D. G. Nocera, and Y. S. Lee, "Topological Magnon Bands in a Kagome Lattice Ferromagnet", Phys. Rev. Lett. {\bf 115}, 147201 (2015).

\bibitem{Lee02}  S. H. Lee, C. Broholm, W. Ratcliff, G. Gasparovic, Q. Huang, T. H. Kim, and S. W. Cheong, ''Emergent excitations in a geometrically frustrated magnet'', Nature {\bf 418}, 856 (2002).

\bibitem{Guguchia20} Z. Guguchia, J. A. T. Verezhak, D. J. Gawryluk, S. S. Tsirkin, J. X. Yin, I. Belopolski, H. Zhou, G. Simutis, S. S. Zhang, T. A. Cochran et al., ''Tunable anomalous Hall conductivity through volume-wise magnetic competition in a topological kagome magnet'', Nat. Commun. {\bf 11}, 559 (2020).
\bibitem{Lachman20} E. Lachman, R. A. Murphy, N. Maksimovic, R. Kealhofer, S. Haley, R. D. McDonald, J. R. Long, and J. G. Analytis, ''Exchange biased anomalous Hall effect driven by frustration in a magnetic kagome lattice'', Nat. Commun. {\bf 11}, 560 (2020).
\bibitem{Zhang21} Q. Zhang, S. Okamoto, G. D. Samolyuk, M. B. Stone, A. I. Kolesnikov, R. Xue, J. Yan, M. A. McGuire, D. Mandrus, and D. A. Tennant, "Unusual Exchange Couplings and Intermediate Temperature Weyl State in ${\mathrm{Co}}_{3}{\mathrm{Sn}}_{2}{\mathrm{S}}_{2}$", Phys. Rev. Lett. {\bf 127}, 117201 (2021).
\bibitem{Lee22} Y. Lee, R. Skomski, X. Wang, P. P. Orth, A. K. Pathak, B. N. Harmon, R. J. McQueeney, I. I. Mazin, Liqin Ke, "Interplay between magnetism and band topology in Kagome magnets RMn$_6$Sn$_6$",  arXiv:2201.11265.


\end{thebibliography}

\begin{thebibliography}{99}

\bibitem{Malaman99} B. Malaman, et al., ''Magnetic properties of RMn$_6$Sn$_6$ (R=Gd-Er) compounds from neutron diffraction and M{\"o}ssbauer measurements", {\it J. Magn. Magn. Mater.} {\bf 202}, 519-534 (1999).
\bibitem{Riberolles22}  S. X. M. Riberolles, Tyler J. Slade, D. L. Abernathy, G. E. Granroth, Bing Li, Y. Lee, P. C. Canfield, B. G. Ueland, Liqin Ke, R. J. McQueeney, "Low-Temperature Competing Magnetic Energy Scales in the Topological Ferrimagnet TbMn$_6$Sn$_6$", Phys. Rev. X, {\bf 12}, 021043 (2022).
\bibitem{Ma21} W. Ma, X. Xu, J.-X. Yin, H. Yang, H. Zhou, Z.-J. Cheng, Y. Huang, Z. Qu, F. Wang, M. Z. Hasan et al., ``Rare Earth Engineering in $R$Mn$_6$Sn$_6$ ($R=$Gd-Tm,Lu) Topological Kagome Magnets'', Phys. Rev. Lett. {\bf 126}, 246602 (2021).
\bibitem{Methfessel00} M. Methfessel, M. van Schilfgaarde, and R. A. Casali, Chapter 3 a full-potential LMTO method based on smooth Hankel functions, in
Electronic Structure and Physical Properties of Solids: The Uses of the LMTO Method, Lecture Notes in Physics, Vol. 535, edited by
H. Dreysse (Springer-Verlag, Berlin, 2000).
\bibitem{lee2022arxiv} Y. Lee, R. Skomski, X. Wang, P. P. Orth, A. K. Pathak, B. N. Harmon, R. J. McQueeney, I. I. Mazin, and L. Ke, ``Interplay between
magnetism and band topology in Kagome magnets RMn$_6$Sn$_6$", arXiv:2201.11265 (2022).
\bibitem{kotani2008jpcm} T. Kotani and M. van Schilfgaarde, ``Spin wave dispersion based on the quasiparticle self-consistent GW method: NiO, MnO and $\alpha$-MnAs",
J. Phys.: Condens. Matter {\bf 20}, 295214 (2008).
\bibitem{ke2021ncm} L. Ke and M. I. Katsnelson, ``Electron correlation effects on exchange interactions and spin excitations in 2D van der Waals materials", npj
Comp. Mater. {\bf 7}, 1 (2021).
\bibitem{li2021prb} B. Li, D. M. Pajerowski, S. X. M. Riberolles, L. Ke, J.-Q. Yan, and R. J. McQueeney, ``Quasi-two-dimensional ferromagnetism and
anisotropic interlayer couplings in the magnetic topological insulator MnBi$_2$Te$_4$", Phys. Rev. B {\bf 104}, L220402 (2021).
\bibitem{ke2011prbr} L. Ke, M. van Schilfgaarde, J. Pulikkotil, T. Kotani, and V. Antropov, ``Low-energy coherent stoner-like excitations in CaFe$_2$As$_2$", Phys.
Rev. B: Rapid Commun. {\bf 83}, 060404 (2011).
\bibitem{kotani2007prb} T. Kotani, M. van Schilfgaarde, and S. V. Faleev, ``Quasiparticle self-consistent gw method: A basis for the independent-particle approximation",
Phys. Rev. B {\bf 76}, 165106 (2007).
\bibitem{aryasetiawan1994prbA} F. Aryasetiawan and O. Gunnarsson, ``Product-basis method for calculating dielectric matrices", Phys. Rev. B {\bf 49}, 16214 (1994).

\end{thebibliography}
\end{document}